\documentstyle[12pt,epsf,a4,epsfig]{article}

\newcommand{\beq}{\begin{equation}}
\newcommand{\eeq}{\end{equation}}
\newcommand{\beqa}{\begin{eqnarray}}
\newcommand{\eeqa}{\end{eqnarray}}
\newcommand{\no}{\nonumber}
\newcommand{\q}{\quad}
\newcommand{\qq}{\qquad}

\begin{document}

\hfill 

\hfill 

\bigskip\bigskip

\begin{center}

{{\Large\bf Photoproduction of $\eta$ and $\eta'$ mesons off nucleons close to
threshold\footnote{Work supported in part by the DFG}}}

\end{center}

\vspace{.4in}

\begin{center}
{\large B. Borasoy\footnote{email: borasoy@physik.tu-muenchen.de}}

\bigskip

\bigskip

Physik Department\\
Technische Universit{\"a}t M{\"u}nchen\\
D-85747 Garching, Germany \\

\vspace{.2in}

\end{center}

\vspace{.7in}

\thispagestyle{empty} 

\begin{abstract}
The importance of Born terms and resonance exchange for
$\eta$ and $\eta'$ photoproduction off both the
proton and neutron
within $U(3)$ baryon chiral perturbation theory is investigated. 
Low-lying resonances such as
the vector mesons and $J^P= 1/2^+, 1/2^-$ baryon resonances are included
explicitly and their contributions together with the Born terms are
calculated. The coupling constants of the resonances are determined from strong
and radiative decays. We obtain reasonable agreement with experimental data
near threshold.
\end{abstract}

\vfill

\section{Introduction}  
Photoproduction of mesons is a tool to study baryon resonances and the
investigation of transitions between these states provides a crucial test for
hadron models. The dominance of the $\Delta (1232)$ in the photoproduction of
pions, e.g., has allowed to extract information on its electromagnetic
transition amplitudes. 
Because of their hadronic decay modes nucleon resonances have large overlapping
widths, which makes it difficult to study individual states, but selection
rules in certain decay channels can reduce the number of possible resonances.
The isoscalars $\eta$ and $\eta'$ are such examples since, due to isospin
conservation, only the isospin-$\frac{1}{2}$ excited states decay into the
$\eta N$ and $\eta' N$ channels.
In recent years both $\eta$ and $\eta'$ photoproduction have been of
considerable interest. The $\eta$ photoproduction of protons has been measured
at MAMI \cite{Kru} and resonance parameters of the $S_{11} (1535) $
resonance and the
electromagnetic coupling $\gamma p \rightarrow  S_{11}$ 
have been extracted from the data.
On the theoretical side, both an effective Lagrangian approach \cite{BeM} and
coupled channel models \cite{Be,Ti} are used to investigate $\eta$ 
photoproduction in the $S_{11} (1535)$ resonance region.
In these approaches the coupling of the $\eta$ to the nucleons is described by
both a pseudovector and a pseudoscalar term and the coupling constant and the
coupling structure of the Born terms is unknown. In \cite{Ti} it has been 
shown that
differential cross sections are rather sensitive to the assumptions about this
vertex. But within the framework of chiral perturbation theory 
this coupling is fixed at lowest
order by making use of the chiral $SU(3)_L \times SU(3)_R$
symmetry of the Lagrangian, whereas explicitly chiral
symmetry breaking terms appear at higher orders. The $SU(3)_L \times SU(3)_R$
symmetric limit
provides therefore a convenient starting point which overcomes the problem of
fixing the $\eta NN$ vertex.
The $SU(3)$ chiral meson-baryon Lagrangian has been used in a coupled channel
model \cite{KWW} and by adjusting a few parameters a large amount of 
low-energy data was described.
All the above mentioned investigations have in common that they treat the
$\eta$ meson as a pure $SU(3)$ octet state $\eta_8$ and mixing of $\eta_8$ with
the corresponding singlet state  $\eta_0$ which yields the physical states
$\eta$ and  $\eta'$ is neglected.

The  $\eta'$ is interesting by itself. The QCD Lagrangian with massless quarks
exhibits an $SU(3)_L \times SU(3)_R$ chiral symmetry which is broken down
spontaneously to $SU(3)_V$, 
giving rise to a Goldstone boson octet of pseudoscalar mesons
which become massless in the chiral limit of zero quark masses.
On the other hand, the axial $U(1)$ symmetry of the QCD Lagrangian is broken by
the anomaly. 
The corresponding pseudoscalar singlet would otherwise have a mass comparable
to the pion mass \cite{W}. Such a particle is missing in the spectrum and the
lightest candidate would be the $\eta'$ with a mass of 958 MeV which is
considerably heavier than the octet states.
In conventional chiral perturbation theory the $\eta'$ 
is not included explicitly, although it does show up in the form of a
contribution to a coupling coefficient of the Lagrangian, a so-called
low-energy constant (LEC).

The   $\eta'$ photoproduction has been investigated 
theoretically both in \cite{ZMB} and
\cite{Li}. In the effective Lagrangian approach of \cite{ZMB} a pseudoscalar 
coupling
of the $\eta'$ to the nucleons was chosen and it was concluded that the
$\eta' N$ decay channel is dominated by the not so well established 
$D_{13} (2080)$
resonance, whereas in the quark model used in \cite{Li} the off-shell 
effects of
the $S_{11} (1535)$ were prominent. 
In contrast, the experimental data from ELSA
\cite{Pl} suggested the coherent excitation of two resonances 
$S_{11} (1897)$ and $P_{11} (1986)$.
Recently, the $\eta'$ has been included in baryon chiral perturbation theory 
in a systematic fashion. Within
this framework it is possible to calculate the importance of the 
$S_{11} (1535)$
resonance for the $\eta' N$ decay channel without making any assumptions on the
$\eta' NN$ coupling.

The aim of this paper is to clarify the role of Born terms
and low-lying resonances for both 
$\eta$ and $\eta'$ photoproduction within the framework of baryon chiral
perturbation theory.
The simultaneous description of $\eta$ and $\eta'$ photoproduction will also
allow the inclusion of the $\eta$-$\eta'$ mixing in a rigorous way.
We restrict ourselves to the threshold region and to the calculation of Born
terms and the leading resonances both in the $t$- channel and in the $s$- 
and $u$-
channels. Such a simplified treatment of $\eta$ and $\eta'$ photoproduction
will not allow us to reproduce all the experimental data in detail. Here, we
are rather concerned with qualitative agreement and a rough estimate of the
importance of low-lying resonances for both decay channels. In order to obtain
a better description of the experimental data, one has to include chiral loops
and further resonances, but this is beyond the scope of the present
investigation.
This work should therefore be considered to be mainly a check if the inclusion
of $\eta$ and $\eta'$ mesons in a nonet of pseudoscalar mesons as proposed in
\cite{B} leads to an adequate description for processes of $\eta$
and $\eta'$  mesons with baryons.

The $\eta$ photoproduction off the neutron which has been measured at ELSA
\cite{Ho} provides a further test of our simple model and we are able to give
predictions for the cross sections for $\eta'$ photoproduction off the neutron.
In the next section, we present the necessary formalism for photoproduction
of $\eta$ and $\eta'$ mesons. The effective chiral Lagrangian including
explicitly low-lying resonances is given in Sec. 3. 
The invariant amplitudes are
shown in Sec. 4 and Sec. 5 contains the numerical results. We conclude with a
summary in Sec. 6. 
The determination of the baryon resonance couplings is relegated to the 
Appendix.

\section{General Formalism}
The $T$-matrix element for the processes $N(p_1) + \gamma(k) \rightarrow
N(p_2) + \phi(q)$ with $\phi = \eta$ or $\eta^{\prime}$ is given by
\beq
\langle  p_2, q \; \mbox{out} | p_1, k \; \mbox{in} \rangle  =
\delta_{fi} + ( 2 \pi)^4 i \delta^{(4)}(p_2 + q - p_1 - k) T_{fi} .
\eeq
The Mandelstam variables are
\beqa
s &=& (k + p_1)^2 = (q + p_2)^2 \no \\
t &=& (k -q)^2 = (p_1 - p_2)^2 \no \\
u &=& (k - p_2)^2 = (q - p_1)^2
\eeqa
subject to the constraint $ s+t+u= 2 M_N^2 + m_\phi^2$ with $M_N$ and
$m_\phi$ being the mass of the nucleon and the pseudoscalar meson,
respectively. The invariant four-momentum transfer squared, $t$, can be related
to the scattering angle $\vartheta$ in the c.m. system via
\beq
t = m^2_\phi - 2 q^0 k^0 + 2 |{\bf q}| |{\bf k}| z
\eeq
with $z = \cos \vartheta$.
The equivalent photon energy in the laboratory system is given by
\beq
E_\gamma = \frac{s - M_N^2}{2 M_N},
\eeq
and the threshold values of $s, t$ and $E_\gamma$ are
\beqa
s_{th}  &=&  ( M_N + m_\phi)^2  \no \\
t_{th}  &=&  - \frac{m_\phi^2}{1 + \frac{m_{\phi}}{M_N}}\no \\
E_{\gamma, th}  &=&  m_\phi( 1 + \frac{m_{\phi}}{2 M_N}) .
\eeqa
In the c.m. system, the differential cross section is related to $T$ via
\beqa  \label{dsigma}
\frac{d \sigma}{ d \Omega_{cm}}  &=&  \frac{1}{64 \pi^2 s}  \,
    \frac{\lambda^{1/2} ( s, M_N^2, m_\phi^2)}{s - M_N^2} |T_{fi}|^2
  \no \\
&=&  \frac{1}{64 \pi^2 s} \frac{|{\bf q}|}{|{\bf k}|} |T_{fi}|^2
\eeqa
with
\beq
\lambda ( s, M_N^2, m_\phi^2) = [ s - ( M_N + m_\phi)^2] \,
    [ s - ( M_N - m_\phi)^2] ,
\eeq
and our normalization is such that $\bar{u}{u} = 2 M_N$.
In general, $T$ can be decomposed as
\beq
T_{fi} = i \epsilon_\mu \bar{u}_2 \sum_{i =1}^{8} B_i {\cal N}_i^\mu u_1
\eeq
with the invariant amplitudes
\beqa
{\cal N}^1_\mu & =& \gamma_5 \gamma_\mu k \! \! \! / ,  \qq 
{\cal N}^2_\mu = 2 \gamma_5 P_\mu , \qq 
{\cal N}^3_\mu = 2 \gamma_5 q_\mu , \no \\
{\cal N}^5_\mu & =& \gamma_5 \gamma_\mu,  \qq 
{\cal N}^6_\mu = \gamma_5 k \! \! \! /  P_\mu , \qq 
{\cal N}^8_\mu = \gamma_5 k \! \! \! /  q_\mu , 
\eeqa
where we have neglected the operators $B_4$ and $B_7$ since they vanish for
real photons and we use $P = \frac{1}{2}( p_1 + p_2)$.
From current conservation one obtains the relations
\beqa
B_3 &=& - \frac{1}{2} \, \frac{s-u}{s+u - 2 M_N^2} \, B_2  \no \\
B_5 &=& - \frac{1}{4} (s-u) B_6 - \frac{1}{2} \Big[ s+u - 2 M_N^2 \Big] B_8 .
\eeqa
It is therefore more convenient to define a set of independent amplitudes
\cite{CGLN}
\beq
T_{fi} = i \bar{u}_2 \sum_{i =1}^{4} A_i {\cal M}_i u_1
\eeq
with
\beqa
{\cal M}_1 &=& \frac{1}{2} \gamma_5 \gamma_\mu \gamma_\nu F^{\mu \nu}, \qq
{\cal M}_2 = 2 \gamma_5 P_\mu q_\nu F^{\mu \nu}, \no \\
{\cal M}_3 &=& \gamma_5 \gamma_\mu q_\nu F^{\mu \nu}, \qq
{\cal M}_4  = 2 \gamma_5 \gamma_\mu P_\nu F^{\mu \nu} - 2M_N {\cal M}_1  
\eeqa
and $F_{\mu \nu} = \epsilon_\mu k_\nu - \epsilon_\nu k_\mu$.
The $A_i$ obey the crossing relations
\beqa
A_i(s,u) &=&  A_i(u,s)   \qq  i = 1,2,4 \no \\
A_3(s,u) &=&  -A_3(u,s) 
\eeqa
and are related to the $B_i$ via
\beqa
A_1 &=& B_1 - M_N B_6 , \qq \q
A_2 = \frac{2}{s+u - 2 M_N^2} \, B_2 ,  \no \\
A_3 &=& -B_8 , \qq \qq \q
A_4 = - \frac{1}{2} B_6 .
\eeqa
The differential cross section can be written in terms of products of the $A_i$
with their complex conjugates $A_i^*$
\beq
\frac{d \sigma}{ d \Omega_{cm}} = \frac{1}{64 \pi^2 s}  \,
    \frac{\lambda^{1/2} ( s, M_N^2, m_\phi^2)}{s - M_N^2} \sum_{i,j = 1}^4 y_{i
    j} A_i A_j^* .
\eeq
The coefficients $y_{ij} = y_{ji}$ read
\beqa
y_{11} &=& [ s - M_N^2] [ M_N^2 - u] \no \\
y_{12} &=& \frac{1}{2} \Big[ t( su - M_N^4 +M_N^2 m_\phi^2) -  
                       M_N^2 m_\phi^4 \Big] \no \\
y_{13} &=& \frac{M_N}{2} [ s-u] [ m_\phi^2 -t] \no \\
y_{14} &=& \frac{M_N}{2}[ m_\phi^2 -t]^2 \no \\
y_{22} &=& - t y_{12} \no \\
y_{23} &=& y_{24} = 0  \no \\
y_{33} &=& \frac{t}{4} [ 2 M_N^4 - 2 M_N^2 m_\phi^2 - s^2 - u^2] + 
           \frac{1}{2}  M_N^2 m_\phi^4 \no \\
y_{34} &=& - \frac{t}{2 M_N} y_{13} \no \\
y_{44} &=&  y_{33} - M_N^2 [ m_\phi^2 -t]^2 .
\eeqa
The total cross section is obtained by integrating over the angular variable
$z = \cos \vartheta$
\beq
\sigma_{tot}(s) = 2 \pi \int_{-1}^{1} dz \frac{d \sigma}{ d \Omega_{cm}} (s,z).
\eeq
For the multipole decomposition one expresses the transition amplitude in terms
of Pauli spinors and matrices
\beq
\frac{1}{8 \pi \sqrt{s}} i \bar{u}_2 \sum_{i =1}^{4} A_i {\cal M}_i u_1=
\chi^\dagger_2 {\cal F} \chi_1 .
\eeq
The most general form for ${\cal F}$ reads
\beq
{\cal F} = i \mbox{\boldmath$ \sigma$} \cdot 
      \mbox{\boldmath$\epsilon$} {\cal F}_1 +
   \mbox{\boldmath$\sigma$} \cdot {\bf \hat{q}} \, 
     \mbox{\boldmath$ \sigma$} \cdot ( {\bf \hat{k}} \times
   \mbox{\boldmath$ \epsilon$} ) \, {\cal F}_2 + i \mbox{\boldmath$ \sigma$} 
    \cdot {\bf \hat{k}} \,
   {\bf \hat{q}} \cdot \mbox{\boldmath$ \epsilon$} \, {\cal F}_3 + 
   i \mbox{\boldmath$ \sigma$} \cdot {\bf \hat{q}} \, {\bf \hat{q}} 
     \cdot \mbox{\boldmath$ \epsilon$} \, {\cal F}_4 
\eeq
and the ${\cal F}_i$ are related to the $A_i$ via
\beqa
{\cal F}_1  &=&  (\sqrt{s} - M_N) \, \frac{N_1 N_2}{8 \pi \sqrt{s}}
   \,    \Big[ A_1 + \frac{k \cdot q}{\sqrt{s} - M_N} A_3 + \Big(
      \sqrt{s} - M_N - \frac{k \cdot q}{\sqrt{s} - M_N} \Big) A_4 \Big]  \no \\
{\cal F}_2  &=&  (\sqrt{s} - M_N) \, \frac{N_1 N_2}{8 \pi \sqrt{s}} \, 
  \frac{|{\bf q}|}{E_2 + M_N} \no \\
&&  \qq \qq \times \Big[-A_1 + \frac{k \cdot q}{\sqrt{s} + M_N} 
 A_3 + \Big(\sqrt{s} + M_N - \frac{k \cdot q}{\sqrt{s} + M_N} \Big) A_4 
      \Big]  \no \\
{\cal F}_3  &=&  (\sqrt{s} - M_N) \, \frac{N_1 N_2}{8 \pi \sqrt{s}} \, 
   |{\bf q}| \,  [ -  (\sqrt{s} - M_N) A_2 + A_3 - A_4 ]  \no \\
 {\cal F}_4  &=&  (\sqrt{s} - M_N) \, \frac{N_1 N_2}{8 \pi \sqrt{s}}\, 
  \frac{|{\bf q}|^2}{E_2 + M_N}\,  [ (\sqrt{s} + M_N) A_2 + A_3 - A_4 ] 
\eeqa
with
\beq
N_i = \sqrt{M_N + E_i}, \qq  E_i = \sqrt{M_N^2 + {\bf p}_i^2} .
\eeq
The projection matrix for the lowest multipoles $E_{0+}, M_{1+}, M_{1-}$ and
$E_{1+}$ is given by
\beq
\left( \begin{array}{l}  E_{0+} \\[0.1cm]  M_{1+} \\[0.1cm]  M_{1-} \\[0.1cm]
   E_{1+} \end{array}
\right) =   \int_{-1}^1 d z
\left( \begin{array}{cccc}
\frac{1}{2} P_0 & - \frac{1}{2} P_1 & 0 & \frac{1}{6} [P_0 - P_2] \\[0.1cm]
\frac{1}{4} P_1 & - \frac{1}{4} P_2 & \frac{1}{12} [P_2 - P_0] & 0 \\[0.1cm]
- \frac{1}{2} P_1 & \frac{1}{2} P_0 & \frac{1}{6} [P_0 - P_2] & 0 \\[0.1cm]
\frac{1}{4} P_1 & - \frac{1}{4} P_2 & \frac{1}{12} [P_0 - P_2] & 
\frac{1}{10} [P_1 - P_3] \end{array}
\right)
\left( \begin{array}{l}  {\cal F}_1 \\[0.1cm] {\cal F}_2 \\[0.1cm]
    {\cal F}_3 \\[0.1cm]  {\cal F}_4 
\end{array} \right)
\eeq
with $P_i$ being the Legendre polynomials.
Neglecting higher partial waves the differential cross section is given in
terms of $E_{0+}, M_{1+}, M_{1-}$ and $E_{1+}$ by
\beq  \label{coeff}
\frac{|{\bf k}|}{|{\bf q}|} \frac{d \sigma}{ d \Omega_{cm}}  =  
A + B \cos \vartheta + C \cos^2 \vartheta ,
\eeq
with
\beqa
A &=& |E_{0+}|^2 + |r|^2  \no \\
B &=& 2 \mbox{Re} ( E_{0+} M_P^*) \no \\
C  &=& |M_P|^2 - |r|^2  \no \\
M_P  &=&  3 E_{1+} + M_{1+} - M_{1-} \no \\
2 |r|^2 &=&  |3 E_{1+} - M_{1+} + M_{1-}|^2 + | 2 M_{1+} + M_{1-}|^2 .
\eeqa
This completes the necessary formalism for the processes considered in this
paper.

\section{The effective Lagrangian}
In this section, we will introduce the effective Lagrangian with $\eta$ and
$\eta'$ coupled both to the ground state baryon octet and low-lying
resonances in the $s,u$- and $t$-channel. Recently, a systematic framework  for
the $\eta'$ in baryon chiral perturbation theory has been developed
\cite{B}. 
Here, we will extend this formalism by including explicitly the low-lying
meson and baryon resonances.

Our starting point is the $U(3)_L \times U(3)_R$ chiral effective Lagrangian of
the pseudoscalar meson nonet $(\pi,K,\eta_8,\eta_0)$ 
coupled to the ground state
baryon octet $(N,\Lambda, \Sigma, \Xi)$ at lowest order in the derivative
expansion
\beq
{\cal L} = {\cal L}_\phi + {\cal L}_{\phi B}
\eeq
with
\beq  \label{mes}
{\cal L}_\phi = 
- v_0 \eta_0^2 + \frac{F_\pi^2}{4}  \langle u_{\mu} u^{\mu} \rangle 
+ \frac{F_\pi^2}{4} \langle \chi_+ \rangle + 
   i F_0 v_3 \eta_0 \langle \chi_- \rangle
+ \frac{1}{12} (F_0^2 - F_\pi^2)  
   \langle u_{\mu} \rangle \langle u^{\mu} \rangle 
\eeq
and
\beqa  \label{bar}
{\cal L}_{\phi B} &=& i \langle \bar{B}  \gamma_{\mu}  [D^{\mu},B] \rangle 
 - M_N \langle \bar{B}B \rangle - \frac{1}{2} D \langle \bar{B} \gamma_{\mu}
 \gamma_5 \{u^{\mu},B\} \rangle  \no \\
&& - \frac{1}{2} F \langle \bar{B} \gamma_{\mu} \gamma_5 [u^{\mu},B] \rangle 
- \lambda \langle \bar{B} \gamma_{\mu} \gamma_5 B \rangle 
  \langle u^{\mu} \rangle .
\eeqa
The pseudoscalar meson nonet is summarized in a matrix valued 
field $U(x)$
\begin{equation}
 U(\phi,\eta_0) = u^2 (\phi,\eta_0) = 
\exp \lbrace 2 i \phi / F_\pi + i \sqrt{\frac{2}{3}} \eta_0/ F_0 \rbrace  
, 
\end{equation}
where $F_\pi \simeq 92.4$ MeV is the pion decay constant and the singlet
$\eta_0$ couples to the singlet axial current with strength $F_0$.
The unimodular part of the field $U(x)$ contains the degrees of freedom of
the Goldstone boson octet $\phi$
\begin{eqnarray}
 \phi =  \frac{1}{\sqrt{2}}  \left(
\matrix { {1\over \sqrt 2} \pi^0 + {1 \over \sqrt 6} \eta_8
&\pi^+ &K^+ \nonumber \\
\pi^-
        & -{1\over \sqrt 2} \pi^0 + {1 \over \sqrt 6} \eta_8 & K^0
        \nonumber \\
K^-
        &  \bar{K^0}&- {2 \over \sqrt 6} \eta_8  \nonumber \\} 
\!\!\!\!\!\!\!\!\!\!\!\!\!\!\! \right) \, \, \, \, \, ,  
\end{eqnarray}
while the phase det$U(x)=e^{i\sqrt{6}\eta_0/F_0}$
describes the $\eta_0$.\footnote{For details the reader is referred to
\cite{B}.} 
In order to incorporate the baryons into the effective theory it is convenient
to form an object of axial-vector type with one derivative
\beq
u_{\mu} = i u^\dagger \nabla_{\mu} U u^\dagger
\eeq
with $\nabla_{\mu}$ being the covariant derivative of $U$.
The expression $\langle \ldots \rangle$ denotes the trace in flavor space
and the quark mass matrix ${\cal M} = \mbox{diag}(m_u,m_d,m_s)$
enters in the combinations
\beq
\chi_\pm = 2 B_0 ( u {\cal M} u \pm  u^\dagger {\cal M} u^\dagger)
\eeq
with $B_0 = - \langle  0 | \bar{q} q | 0\rangle/ F_\pi^2$ the order
parameter of the spontaneous symmetry violation.
Expanding the Lagrangian ${\cal L}_\phi$ in terms of the meson fields
one observes terms quadratic in the meson fields that contain the factor
$\eta_0 \eta_8$ which leads to $\eta_0$-$\eta_8$ mixing.
Such terms arise from the explicitly symmetry breaking terms 
$\frac{F_\pi^2}{4} \langle \chi_+ \rangle + i F_0 v_3 \eta_0 
\langle \chi_- \rangle$ and read
\beq
- \Big( \frac{2 \sqrt{2}}{3} \frac{F_\pi}{F_0} + \frac{8}{\sqrt{3}} 
\frac{F_0}{F_\pi} v_3 \Big) B_0 (\hat{m} -m_s) \eta_0 \eta_8
\eeq
with $\hat{m} = \frac{1}{2} (m_u + m_d)$.
The states $\eta_0$ and $\eta_8$ are therefore not mass eigenstates.
The mixing yields the eigenstates $\eta$ and $\eta'$,
\beqa
| \eta \rangle & = & \cos \theta \, | \eta_8 \rangle -
                     \sin \theta \, | \eta_0 \rangle  \no \\
| \eta'\rangle & = & \sin \theta \, | \eta_8 \rangle +
                     \cos \theta \, | \eta_0 \rangle , 
\eeqa
which is valid in the leading order of flavor symmetry breaking and
we have neglected other pseudoscalar isoscalar states which could mix
with both $\eta_0$ and $\eta_8$.
The $\eta$-$\eta'$ mixing angle can be determined from the two
photon decays of $\pi^0, \eta, \eta'$, which require a mixing angle
around -20$^\circ$ \cite{exp}. 
We will make use of this experimental input in order to
diagonalize the mass terms of the effective mesonic Lagrangian.

The baryonic Lagrangian consists of the free kinetic term and the axial-vector
couplings of the mesons to the baryons. The values of the LECs $D$ and $F$ can
be extracted from semileptonic hyperon decays.
A fit to the experimental data delivers 
$D=0.80 \pm 0.01$ and $F=0.46 \pm 0.01$ \cite{CR}.
We leave the third axial-vector coupling $\lambda$ undetermined for the time
being.
The covariant derivative of the baryon field is given by
\beq
[ D_\mu, B] = \partial_\mu B + [ \Gamma_\mu, B]
\eeq
with the chiral connection
\beq
\Gamma_\mu \simeq  - i v_\mu =  i e Q {\cal A}_\mu
\eeq
to the order we are working and $Q = \frac{1}{3} \mbox{diag}(2,-1,-1)$
is the quark charge matrix.
Note that there is no pseudoscalar coupling of $\eta_0$ to the baryons of
the form $\eta_0 \bar{B}  \gamma_5 B$. Such a term is in principle possible but
can be absorbed by the $\lambda$-term in Eq. (\ref{bar}) 
by means of the equation of motion for the baryons. 
Since we confine ourselves to the lowest order Lagrangian for the ground state
baryon octet, higher order terms are omitted. The main purpose here is to
investigate the importance of lowest order Born terms and resonance
exchange. In order to obtain an improved description one must, of course,
include higher order counterterms such as the anomalous magnetic moments of the
nucleons.

We now proceed by including explicitly low-lying resonances in our theory.
In the $t$-channel the lowest-lying resonances are the nonet of the
vector mesons $(\rho, K^*, \omega_8, \omega_0)$. 
According to \cite{Pl} no evidence
for $\rho$ and $\omega$ exchange in the $t$-channel is given. We will
investigate this statement within our framework.
The vector mesons couple in a chiral invariant way to the baryons and the
pseudoscalar mesons via the Lagrangian
\beq
{\cal L} = {\cal L}_{VBB} + {\cal L}_{V \gamma \phi}
\eeq
with
\beq  \label{vnn}
{\cal L}_{VBB} = g^D_{VBB} \langle \bar{B}  \Gamma_{\mu} \{V^{\mu},B\} \rangle 
 + g^F_{VBB} \langle \bar{B}  \Gamma_{\mu}  [V^{\mu},B] \rangle +
   g^0_{VBB} \omega_0^\mu \langle \bar{B}  \Gamma_{\mu} B \rangle ,
\eeq
where the operator $\Gamma_{\mu}$ involves a vector and a tensor coupling
\beq
\Gamma_{\mu} = \gamma_{\mu} + i \frac{\kappa_V}{2 M_N} \sigma_{\mu \nu}
                 ( p'-p)^\nu .
\eeq
The electromagnetic piece of the Lagrangian is given by
\beq  \label{elec}
{\cal L}_{V \gamma \phi} = -g^8_{V \gamma \phi} \epsilon^{\mu \nu \alpha \beta}
\langle V_\mu \{ u_\nu, F_{\alpha \beta}^+\} \rangle   - g^0_{V \gamma \phi}
\epsilon^{\mu \nu \alpha \beta} \omega^0_\mu \langle u_\nu F_{\alpha \beta}^+
\rangle .
\eeq
The quantity $F_{\alpha \beta}^+$ contains the electromagnetic field strength
tensor $F_{\alpha \beta}$ of $v_\mu$
\beqa
F_{\alpha \beta}^+  &\equiv&   u^\dagger F_{\alpha \beta} u  +
          u F_{\alpha \beta} u^\dagger  \no \\
  &=&   2 ( \partial_\alpha v_\beta - \partial_\beta v_\alpha ) + {\cal
  O}(\phi^2)  .
\eeqa
The states $\omega_8$ and $\omega_0$ mix to yield the physical states $\omega$
and $\phi$. The mixing is characterized by the angle $\varphi$
\beqa
| \omega \rangle & = & \cos \varphi \, | \omega_8 \rangle -
                     \sin \varphi \, | \omega_0 \rangle  \no \\
| \phi \rangle & = & \sin \varphi \, | \omega_8 \rangle +
                     \cos \varphi \, | \omega_0 \rangle 
\eeqa
with $\varphi \simeq 40^\circ$ \cite{Du}.
The LECs of the effective Lagrangian are usually expressed in terms of the
physical couplings $g_{VN}$ of $\rho_0, \omega$ and $\phi$ to the proton
\beq
{\cal L}_{Vpp} = \frac{1}{2} \bar{p} \Gamma_\mu p \sum_{V= \rho_0, \omega,\phi}
   g_{VN} V^\mu .
\eeq
Comparison with the Lagrangian in Eq. (\ref{vnn}) leads to
\beqa
g^D_{VBB}  &=&  \frac{ \sqrt{3}}{4 \sqrt{2}}  \Big[ \sqrt{3} g_{\rho N}
  - \cos \varphi \, g_{\omega N}  - \sin \varphi \, g_{\phi N} \Big]  \no \\
g^F_{VBB}  &=&  \frac{ \sqrt{3}}{4 \sqrt{2}}  \Big[\frac{1}{\sqrt{3}} g_{\rho
N} + \cos \varphi \, g_{\omega N} + \sin \varphi \, g_{\phi N} \Big]  \no \\
g^0_{VBB}  &=&  \frac{1}{2} \Big[ \cos \varphi \, g_{\phi N} - \sin \varphi \,
g_{\omega N} \Big] .
\eeqa
It follows immediately, that the coupling of $\rho_0, \omega$ and $\phi$ to the
neutron is given by 
\beq
{\cal L}_{Vnn} = \frac{1}{2} \bar{n} \Gamma_\mu n  \Big(
   - g_{\rho N} \rho_0^\mu + g_{\omega N} \omega^\mu + g_{\phi N} \phi^\mu 
   \Big).
\eeq
The couplings $g_{VN}$ are quite well known, we use $g_{\rho N} = 6.08$ and 
$g_{\omega N} = 3 g_{\rho N}$ \cite{Ba}.
The $g_{\phi N}$ coupling turns out to be much smaller than $g_{\omega N}$
\cite{Du} (in agreement with the OZI suppression),  so we can safely neglect
the $\phi$ meson in our calculations by setting
$g_{\phi N} = 0$. Furthermore, the tensor coupling for the $\rho$ meson is
given by $\kappa_\rho =6$, whereas $\kappa_\omega \simeq 0$. Instead of using a
common tensor coupling $\kappa_V$ for $\rho$ and $\omega$, as prescribed by the
Lagrangian in Eq. (\ref{vnn}), we prefer to work with the physical values
$\kappa_\rho =6$ and $\kappa_\omega = 0$.

We proceed in a similar manner with the electromagnetic piece of the Lagrangian
by defining the physical couplings
\beq
{\cal L}_{V \gamma \phi} =  e \, \epsilon^{\mu \nu \alpha \beta}
    \partial_\alpha {\cal A}_\beta \Big( \partial_\nu \eta 
  \sum_{V= \rho_0, \omega,\phi} g_{V \gamma \eta} V_\mu
 +  \partial_\nu \eta' \sum_{V= \rho_0, \omega,\phi} g_{V \gamma \eta'} V_\mu
 \Big)
\eeq
which are related to the $SU(3)$ couplings via
\beqa
g_{V \gamma \phi}^8 &=&   \Big[ \frac{8}{\sqrt{3} F_0}
\sin \theta - 4 \sqrt{\frac{2}{3}} \frac{1}{F_\pi} \cos \theta \Big]^{-1} 
  g_{\rho \gamma \eta} \no \\
g_{V \gamma \phi}^0 &=&  \frac{\sqrt{3} F_\pi}{4 \cos \theta} \Big[ \sin
\varphi \, g_{\omega \gamma \eta} - \cos \varphi \, 
  g_{\phi \gamma \eta} \Big] .
\eeqa
The experimental values for the $g_{V \gamma \eta}$ can be extracted from the
decay width of radiative decays of the vector mesons
\beq
\Gamma( V \rightarrow \eta \gamma) = \frac{e^2}{96 \pi} g_{V \gamma \eta}^2
    ( m_V - \frac{m_\eta^2}{m_V} )^3 .
\eeq
Using the values for $\Gamma( \rho \rightarrow \eta \gamma)$ and  
$\Gamma( \omega \rightarrow \eta \gamma)$ from \cite{exp} we obtain
\beqa
g_{ \rho \gamma \eta}  &=&  1.8   \; \mbox{GeV}^{-1} \no \\ 
g_{ \omega \gamma \eta}  &=&  0.23   \; \mbox{GeV}^{-1}  .
\eeqa
Once the couplings of the vector mesons with $\eta$ have been determined,
their coupling strength to $\eta'$ could in principle be calculated by making
use of the Lagrangian in Eq. (\ref{elec}).
But in order to get a more accurate estimate of these couplings, it is
preferable to extract the coupling constants $g_{ \rho \gamma \eta'}$ and $g_{
\omega \gamma \eta'}$ directly from the decay widths of the pertinent radiative
decays
\beq
\Gamma( \eta' \rightarrow V \gamma) = \frac{e^2}{32 \pi} g_{V \gamma \eta'}^2
    ( m_{\eta'} - \frac{m_V^2}{m_{\eta'}} )^3 .
\eeq
We obtain
\beqa
g_{ \rho \gamma \eta'}  &=&  1.31   \; \mbox{GeV}^{-1} \no \\ 
g_{ \omega \gamma \eta'}  &=&  0.45  \; \mbox{GeV}^{-1}  .
\eeqa
This determines completely the contributions of the vector mesons.
Note that the vector meson contribution is usually reduced, e.g., 
by using a form factor \cite{ZMB}. However,
this effect should be reasonably small for $\eta$ and $\eta'$ 
photoproduction close to threshold.

Baryon resonances contribute in the $s$- and $u$-channel. In this work we
consider the lowest-lying $S$- and $P$-wave baryon resonances, i.e. the $J^P =
1/2^+$ and $1/2^-$ octets which include $P_{11}(1440)$ and
$S_{11}(1535)$, respectively. We will neglect higher partial waves 
baryon resonances such as $D_{13}(1520)$. Both $P_{11}(1440)$ and
$S_{11}(1535)$ contribute to $\eta$ photoproduction and experimentally
dominance of $S_{11}(1535)$ is found \cite{Kru}.
On the other hand, the situation for $\eta'$ photoproduction is not so
clear. It has been discussed in the literature whether the data for 
$\eta'$ photoproduction on the proton can be understood by taking resonances in
the 2 GeV region or off-shell effects of $S_{11}(1535)$ into account
\cite{ZMB, Li}.
In the present investigation we will extend the formalism for the chiral $U(3)$
Lagrangian presented in \cite{B} by including the $J^P = 1/2^+$ and 
$1/2^-$ baryon resonance octets.
As we will see shortly, the parameters of such a Lagrangian can be fixed from
strong and radiative decays of the resonances. This allows us to predict the
contributions of $P_{11}(1440)$ and $S_{11}(1535)$ both for $\eta$ and $\eta'$ 
photoproduction. Within our effective Lagrangian approach the results will be
predictive and no additional parameters have to be fitted to obtain agreement
with experiment.
Therefore, the application of the $U(3)$ formalism within baryon chiral
perturbation theory to photoproduction processes is a highly nontrivial check
of this method. It can also clarify the importance of the $S_{11}(1535)$ 
resonance for $\eta'$ photoproduction.
To this end, it is sufficient to obtain a rough estimate for the resonance
contributions. In order to achieve better agreement with experiment, one has to
consider further resonances, e.g. $D_{13}(1520)$ and $S_{11}(1650)$, and
include chiral loop corrections. But this is beyond the scope of the present
investigation and the calculations are performed at tree level.

Let us first consider the spin-$1/2^+$ octet which we denote by $P$. The
octet consists of $N^*(1440), \Sigma^*(1660), \Lambda^*(1600), \Xi^*(?)$
and the effective Lagrangian of the $P$-wave octet coupled to the ground state
baryon octet takes the form
\beq
{\cal L} = {\cal L}_{P} + {\cal L}_{\phi B P}
\eeq
with the kinetic term
\beq
{\cal L}_{P} = 
i \langle \bar{P}  \gamma_{\mu}  [D^{\mu},P] \rangle 
 - M_P \langle \bar{P}P \rangle .
\eeq
Since for the processes considered here only $N^*(1440)$ contributes, we set
$M_P = 1.44$ GeV.
The interaction terms of the $P$-wave resonances with the ground state baryon
octet read
\beqa  \label{res1}
{\cal L}_{\phi B P} &=& - \frac{1}{2} D_P \langle \bar{P} \gamma_{\mu}
 \gamma_5 \{u^{\mu},B\} \rangle  
 - \frac{1}{2} F_P \langle \bar{P} \gamma_{\mu} \gamma_5 [u^{\mu},B] 
 \rangle - \lambda_P \langle \bar{P} \gamma_{\mu} \gamma_5 B \rangle 
  \langle u^{\mu} \rangle \no \\
&&
+ d_P \langle \bar{P} \sigma^{\mu \nu} \{ F^+_{\mu \nu}, B \} \rangle 
+ f_P \langle \bar{P} \sigma^{\mu \nu} [ F^+_{\mu \nu}, B ] \rangle 
+ \mbox{h.c.} \; .
\eeqa
A possible $\eta_0 \bar{P} \gamma_5 B$ 
term can again be eliminated by using the
equation of motion for baryons.
The coupling constants $D_P, F_P$ and $d_P, f_P$ can be determined from 
strong and radiative decays of the $N^*(1440)$ resonance, cf. App. A; we use
the central values
\beqa
D_P &=& 0.32 ,  \qq  F_P = 0.16  \no \\
d_P &=& -0.05 \; \mbox{GeV}^{-1}  ,  \qq  f_P = 0.08 \; \mbox{GeV}^{-1}   .
\eeqa
The spin-$1/2^-$ octet consists of $N^*(1535), \Lambda^*(1670), \Sigma^*(1750),
\Xi^*(?)$ and the pertinent Lagrangian reads
\beq
{\cal L} = {\cal L}_{S} + {\cal L}_{\phi B S}
\eeq
with the kinetic term
\beq
{\cal L}_{S} = 
i \langle \bar{S}  \gamma_{\mu}  [D^{\mu},S] \rangle 
 - M_S \langle \bar{S}S \rangle 
\eeq
and the interaction part
\beqa  \label{res2}
{\cal L}_{\phi B S} &=& - \frac{i}{2} D_S \langle \bar{S} \gamma_{\mu}
 \{u^{\mu},B\} \rangle  
 - \frac{i}{2} F_S \langle \bar{S} \gamma_{\mu} [u^{\mu},B] 
 \rangle - i \lambda_S \langle \bar{S} \gamma_{\mu} B \rangle 
  \langle u^{\mu} \rangle \no \\
&&
+ i d_S \langle \bar{S} \sigma^{\mu \nu} \gamma_5 
  \{ F^+_{\mu \nu}, B \} \rangle 
+ i f_S \langle \bar{S} \sigma^{\mu \nu} \gamma_5 
   [ F^+_{\mu \nu}, B ] \rangle 
+ \mbox{h.c.} \; .
\eeqa
We set $M_S = 1.535$ GeV and from strong and radiative decays of the $S$-wave
resonances one obtains, cf. App. A,
\beqa
D_S &=& 0.37 ,  \qq  F_S = -0.21 ,  \qq \lambda_S = -0.07   \no \\
d_S &=& -0.07 \; \mbox{GeV}^{-1}  ,  \qq  f_S = -0.06 \; \mbox{GeV}^{-1}   .
\eeqa
Since there exists data on decay channels of the $S$-wave resonances 
into $\eta$,
we are able to fix the coupling $\lambda_S$ by taking $\eta$-$\eta'$ mixing
into account, whereas for $\lambda_P$ we relegate the discussion to Sec. 5.
Several remarks are in order.
First, we would like to point out that our simple ansatz of zero width
resonances will lead to singularities at the resonance mass which
could be circumvented by the use of a finite width.
This will restrict in the case of $\eta$ photoproduction
the validity of our approach to energies very close to threshold
which we are considering in the present work,
whereas it is numerically irrelevant for $\eta'$ photoproduction.
Furthermore, the couplings of the resonances to the nucleons are determined by
their pertinent decay widths. Therefore, their sign is not fixed and a
different choice of their signs leads to changes in our results. We have chosen
the signs of the resonance couplings in such a way, that they lead to better
agreement with existing data for $\eta$ and $\eta'$  photoproduction and do not
present the results for the other possible values of the couplings.
Finally, we have calculated both Born terms using the lowest order chiral
effective Lagrangian and resonance contributions. We would like to emphasize
that this procedure does not imply any double counting. The contributions of
the resonances are hidden only in higher chiral order counterterms
of the effective Lagrangian which we did not take into account in the present
investigation. Born terms like the ones used in this work are not produced by
resonance contributions.

\section{Invariant amplitudes}
We proceed by presenting  the invariant amplitudes for $\eta$ and $\eta'$
photoproduction on the nucleons.
Let us start with the Born terms which vanish in the case of the neutron. The
diagrams contributing to photoproduction off the proton are depicted in
Fig. 1 and they read
\beqa
A_1(p \gamma \rightarrow p \phi)  &=& - 2 M_N e A_\phi
   \Big[ \frac{1}{s - M_N^2} + \frac{1}{u - M_N^2} \Big] \no \\
A_2(p \gamma \rightarrow p \phi)  &=&  4 M_N e A_\phi
   \frac{1}{[s - M_N^2][u - M_N^2]} \no \\
A_3(p \gamma \rightarrow p \phi)  &=& A_4(p \gamma \rightarrow p \phi) =0
\eeqa
with 
\beqa
A_\eta &=& \frac{1}{2 \sqrt{3} F_\pi} [D-3F] \cos \theta + \sqrt{\frac{2}{3}}
  \, \frac{1}{F_0}  [D+3 \lambda] \sin \theta   \no \\
A_{\eta'} &=&\frac{1}{2 \sqrt{3} F_\pi} [D-3F] \sin \theta - \sqrt{\frac{2}{3}}
  \, \frac{1}{F_0}  [D+3 \lambda] \cos \theta   .
\eeqa
This leads to the total cross section
\beqa
\sigma_{tot}(p \gamma \rightarrow p \phi) &=&  - \frac{\lambda^{1/2} 
  ( s, M_N^2, m_\phi^2) }{4 \pi s [s-M_N^2]}  M_N^2 e^2 A_\phi^2 \, \Bigg(\, 
 \frac{1}{2s} [3s - M_N^2 + m_\phi^2] \no \\
&& - \frac{4s m_\phi^2}{[s-M_N^2]^2} -
 \frac{4s}{\Delta_\phi} \Big[ 
 \frac{1}{2} - m_\phi^2 \frac{s + M_N^2 - m_\phi^2}{[s-M_N^2]^2} \Big] \no \\
&&  \times \ln \Big[ \frac{s + M_N^2 - m_\phi^2 + \Delta_\phi }{ 
   2 \sqrt{s} M_N } \Big] \, \Bigg) 
\eeqa
with
\beq
\Delta_\phi = \sqrt{(s - M_N^2 + m_\phi^2)^2 - 4s m_\phi^2}. 
\eeq
Vector meson exchange is shown in Fig. 2. One has to add the
following terms to the invariant amplitudes for photoproduction on the proton
\beqa
A_1(p \gamma \rightarrow p \phi)  &=& \frac{e \kappa_\rho}{4 M_N} g_{\rho N} 
  \, g_{\rho \gamma \phi} \frac{t}{t - M_\rho^2}  \no \\
A_2(p \gamma \rightarrow p \phi)  &=& \frac{e \kappa_\rho}{4 M_N} g_{\rho N} 
  \, g_{\rho \gamma \phi} \frac{1}{t - M_\rho^2}  \no \\
A_3(p \gamma \rightarrow p \phi)  &=&  0 \no \\
A_4(p \gamma \rightarrow p \phi)  &=&  - \frac{e}{2}  
 \sum_{V= \rho_0, \omega,\phi} g_{V N}\,  
  g_{V \gamma \phi} \frac{1}{t - M_V^2} .
\eeqa
For the neutron $g_{\rho N}$ has to be replaced by $-g_{\rho N}$.

We now turn to the baryon resonances. Their contributions are given in Fig. 3
and read for the spin-1/2$^+$ octet in the proton case
\beqa \label{probar} 
A_1(p \gamma \rightarrow p \phi)  &=&  - e \frac{4}{3} (d_P + 3 f_P) P_\phi
\Big[ \frac{u - M_N^2}{u- M_P^2} + \frac{s - M_N^2}{s- M_P^2} \Big] \no \\
A_2(p \gamma \rightarrow p \phi)  &=&  0  \no \\
A_3(p \gamma \rightarrow p \phi)  &=&   e \frac{4}{3} (d_P + 3 f_P) P_\phi
( M_P + M_N) \Big[ \frac{1}{s- M_P^2} - \frac{1}{u- M_P^2} \Big] \no \\
A_4(p \gamma \rightarrow p \phi)  &=&   e \frac{4}{3} (d_P + 3 f_P) P_\phi
( M_P + M_N) \Big[ \frac{1}{s- M_P^2} + \frac{1}{u- M_P^2} \Big] 
\eeqa
with
\beqa
P_\eta &=& \frac{1}{2 \sqrt{3} F_\pi} [D_P-3F_P] \cos \theta + 
\sqrt{\frac{2}{3}}  \, \frac{1}{F_0}  [D_P+3 \lambda_P] \sin \theta   \no \\
P_{\eta'} &=& \frac{1}{2 \sqrt{3} F_\pi} [D_P-3F_P] \sin \theta - 
\sqrt{\frac{2}{3}}  \, \frac{1}{F_0}  [D_P+3 \lambda_P] \cos \theta , 
\eeqa
whereas the results for the neutron are obtained by replacing $d_P + 3 f_P$ by
$-2d_P$ in Eq. (\ref{probar}).
The contributions from the spin-1/2$^-$ resonances read in the case of the
proton 
\beqa \label{barpro}
A_1(p \gamma \rightarrow p \phi)  &=&   e \frac{4}{3} (d_S + 3 f_S) S_\phi
\Big[ \frac{u - M_N^2}{u- M_S^2} + \frac{s - M_N^2}{s- M_S^2} \Big] \no \\
A_2(p \gamma \rightarrow p \phi)  &=&  0  \no \\
A_3(p \gamma \rightarrow p \phi)  &=&   e \frac{4}{3} (d_S + 3 f_S) S_\phi
( M_S - M_N) \Big[ \frac{1}{s- M_S^2} - \frac{1}{u- M_S^2} \Big] \no \\
A_4(p \gamma \rightarrow p \phi)  &=&   e \frac{4}{3} (d_S + 3 f_S) S_\phi
( M_S - M_N) \Big[ \frac{1}{s- M_S^2} + \frac{1}{u- M_S^2} \Big] 
\eeqa
with
\beqa
S_\eta &=& \frac{1}{2 \sqrt{3} F_\pi} [D_S-3F_S] \cos \theta + 
\sqrt{\frac{2}{3}}  \frac{1}{F_0}  [D_S+3 \lambda_S] \sin \theta   \no \\
S_{\eta'} &=& \frac{1}{2 \sqrt{3} F_\pi} [D_S-3F_S] \sin \theta - 
\sqrt{\frac{2}{3}}  \frac{1}{F_0}  [D_S+3 \lambda_S] \cos \theta ,
\eeqa
where for neutrons $d_S + 3 f_S $ in Eq. (\ref{barpro})
has to be replaced by $-2d_S$.

\section{Numerical results}
In this section, we discuss the numerical results for the central values of the
resonance couplings as given in Sec. 3. We were able to determine most LECs by
using experimental data from both semileptonic decays of the ground state
baryon octet and from strong and radiative decays of the baryon resonances. The
coupling constants of the vector meson Lagrangian are quite well known. The
only parameters not fixed so far are the  couplings $\lambda$ and $\lambda_P$
of the axial flavor-singlet baryonic currents. 
We note that from large $N_c$ counting
rules, one expects $|\lambda| < |D|, |F|$ and $|\lambda_P| < |D_P|, |F_P|$.
In order to get an estimate of these couplings we therefore varied their
values within a small range around zero. It turns out that the dependence
on $\lambda_P$ is almost negligible and one can safely set $\lambda_P
=0$. Variation of $\lambda$ leads to some smaller changes in the results and we
find improved agreement with experiment for $\lambda = 0.05$.
Of course, we could achieve better agreement with experiment by fine-tuning all
parameters, but here we are only interested in a rough estimate of the
resonance contributions. For $F_0$ we employ the large $N_c$ identity $F_0 =
F_\pi$. Once the parameters are determined, the resonance
contributions are fixed both for $\eta$ and $\eta'$ photoproduction. This will
particularly clarify the importance of low-lying resonances for  $\eta'$
photoproduction within the chiral $U(3)$ Lagrangian approach presented in this
paper, since our results are predictions rather than a fit to experimental
data. It also serves as a check if the $\eta'$ can in general be included in
baryon chiral perturbation theory in a systematic fashion as proposed in
\cite{B}.

In Fig. 4 we present the total cross sections obtained from our tree level
model. We restrict ourselves to the threshold region, since chiral loop effects
and contributions from further resonances will become more important with
increasing  c.m. energy $s$.
We are able to achieve reasonable agreement with existing data. 
However, the energy bins of the experiment for $\eta'$ photoproduction on the
proton are 100 MeV and 200 MeV wide \cite{Pl} and can therefore not be compared
directly with our theoretical estimates. But we obtain a total cross section
for $\eta'$ photoproduction which is in the same order of magnitude as in the
experiment, i.e. $\simeq 1.5$ $\mu$b.
In particular, 
$\eta'$ photoproduction close to threshold can be understood without taking any
further resonances into account.
The dashed lines in Fig. 4 are the contributions from the Born terms without
the inclusion of resonances. Obviously, the resonances considered in this model
lead to sizeable contributions except for $\eta'$ photoproduction on the
proton in which case the sum of their contributions is almost negligible. 
Furthermore, we are able to reproduce the ratio of $\eta$ photoproduction on
the neutron and the proton, which is experimentally found to be 2/3 \cite{Ho}.
It is also worthwhile to compare our results for
the differential cross sections with experimental data.
In Fig. 5 we show the differential cross sections 
for $\eta$ and $\eta'$ photoproduction on the proton close to threshold.
The data for $\eta'$ photoproduction as shown in Fig. 5b) is the
energy-integrated angular distribution for $1.44$ GeV $ < E_\gamma < 1.54$ GeV.
Dividing by the phase space factor $|{\bf q}|/|{\bf k}| $ in Eq. 
(\ref{dsigma}) in order to account for the wide energy bin 
one obtains experimental data which have a slightly smaller differential cross
section than obtained in our model. This might indicate that we overestimated
the resonances, e.g., by not using a form factor for the vector
mesons. Nevertheless, our simplified treatment of $\eta$ and $\eta'$
photoproduction  shows, that one is able to understand the size of experimental
data just by considering lowest order Born terms and resonance exchange.

In Figs. 6 and 7
we present the results for the multipoles $E_{0+}, M_{1+}, M_{1-}$
and $E_{1+}$  in the threshold region. In order to estimate the significance of
the different resonances, we show in Table 1 the multipoles  for $\eta$ 
photoproduction off the proton at  c.m. energy $s= 708$ MeV both with and
without resonances. We conclude that while the spin-$1/2^+$ resonance octet
does not contribute significantly, both vector mesons and spin-$1/2^-$ 
octet lead to sizeable contributions.
For completeness we present the energy dependence of the coefficients $A, B$
and $C$ from Eq. (\ref{coeff}) in Figs. 8 and 9.

\section{Summary}
In the present work, we studied $\eta$ and $\eta'$ photoproduction on both the
proton and neutron. To this end, we pinned down an effective chiral $U(3)$ 
Lagrangian which describes the interactions of the pseudoscalar meson nonet
$(\pi,K,\eta,\eta')$ with the ground state baryon octet and low-lying
resonances. These include the vector mesons $\rho_0$ and $\omega$ in the
$t$-channel (the $\phi$ meson leads to much smaller contributions and can be
neglected for our purposes), and the $J^P = 1/2^+$ and $1/2^-$ baryon
resonances $P_{11}(1440)$ and $S_{11}(1535)$.
Our aim here is to obtain a rough estimate of the contributions of these
low-lying resonances both for $\eta$ and $\eta'$ photoproduction.
Further resonances have therefore been neglected, such as $S_{11}(1650)$ and
the higher partial wave resonance $D_{13}(1520)$.

Most LECs of the effective Lagrangian can be determined using semileptonic
hyperon decays and both strong and radiative decays of the baryon
resonances. The couplings of the vector mesons are also quite well known. Only
the couplings of the axial flavor-singlet currents of the ground state and
spin-1/2$^+$ resonance baryons, $\lambda$ and $\lambda_P$, could not be fixed
from experiment. Variation of both parameters within a realistic range revealed
that $\lambda_P$ does almost not alter our results and is therefore set to
zero. For $\lambda$ we have chosen a value which leads to improved agreement
with experiment.
Our results are therefore predictions rather than fits to experiment.

We calculated the Born terms and the resonance contributions to $\eta$ and
$\eta'$ photoproduction on the nucleons using the chiral $U(3)$
Lagrangian. Comparison with data close to threshold shows that this simple
model is capable  of producing simultaneously reasonable agreement with 
$\eta$ and $\eta'$ photoproduction both on the neutron and proton, 
i.e. the size of the experimental data for photoproduction of the $\eta'$ meson
can be understood just by taking low-lying resonances into account.
Of course, we do not expect our model to be valid for higher c.m. energies away
from threshold, since other effects such as contributions from further
resonances and chiral loop corrections will become significant.
We are able to reproduce the ratio of $\eta$ photoproduction on the
neutron and the proton, which is experimentally found to be 2/3 \cite{Kru}. 
Reasonable agreement with experiment is also obtained for the 
differential cross sections,
which is almost constant for $\eta$ photoproduction and forward-peaked
in the
case of $\eta'$. Finally, we present results for the multipole
decomposition of $\eta$ photoproduction.
Within our model, we find, that contributions from vector mesons and
$S_{11}(1535)$  dominate, whereas contributions from $P_{11}(1440)$ lead to
some minor corrections.

The present investigation also served as a check if the $\eta'$ meson can be
included in baryon chiral perturbation theory as proposed in \cite{B}.
The findings of the present investigation concerning $\eta'$ photoproduction
can be confirmed, e.g., by employing the chiral $U(3)$ meson-baryon Lagrangian
without explicit resonances in a 
coupled channel approach. This
will generate dynamically the baryon resonances as was shown in \cite{KWW} for 
$\eta$ photoproduction. Work on this subject is currently underway.

\section*{Acknowledgments}
The author wishes to thank N. Kaiser for useful discussions and suggestions.

\appendix 
\def\theequation{\Alph{section}.\arabic{equation}}
\setcounter{equation}{0}
\section{Determination of the baryon resonance couplings } \label{app:a}
In order to determine the strong coupling constants 
of the baryon resonances, we use the decays $N^*(1440) \rightarrow N \pi$,
$\Lambda^*(1600)\rightarrow \Sigma \pi$ and  
$N^*(1535) \rightarrow N \pi$ , $N^*(1535) \rightarrow N \eta$,
$\Lambda^*(1670) \rightarrow \Lambda \eta$, see also \cite{B1}.
The width follows via
\beq
\Gamma = \frac{1}{8 \pi M_R^2} |{\bf q}_\phi| |{\cal T}|^2
\eeq
with
\beq
|{\bf q}_\phi| = \frac{1}{2 M_R} \Big[ (M_R^2-(M_B+m_\phi)^2)
                 \, (M_R^2-(M_B-m_\phi)^2) \Big]^{1/2}
\eeq
being the three--momentum of the meson $\phi = \pi , \eta$
in the rest frame of the resonance.
The terms $M_R$ and $M_B$ are the masses of the resonance and the
ground state baryon, respectively.
We employ the physical masses of the baryons involved in a decay.
The mistake we make in not using  common octet masses
is of higher chiral order and, therefore, beyond the accuracy
of our calculation.
For the transition matrix one obtains in the case of $P$-wave resonances
\beq
|{\cal T}|^2 = \frac{1}{2 F_\pi^2}  ( M_R + M_B)^2 
     \Big[ (M_R - M_B)^2 - m_\phi^2 \Big] A_{P \phi} 
\eeq
and for the $S$-waves
\beq
|{\cal T}|^2 = \frac{1}{2 F_\pi^2}  ( M_R - M_B)^2 
     \Big[ (M_R + M_B)^2 - m_\phi^2 \Big] A_{S \phi}
\eeq
with the coefficients
\beqa
A_{N^*(1440) \, \pi} &=& \frac{3}{2} ( D_P + F_P )^2 , \q
A_{\Lambda^*(1600) \, \pi} = 2 D_P^2 , \q
A_{N^*(1535) \, \pi} = \frac{3}{2} ( D_S + F_S )^2 , \no \\
A_{N^*(1535) \, \eta} &=& \frac{1}{6} \Big[ (D_S - 3 F_S) \cos \theta +
    \sqrt{8} \frac{F_\pi}{F_0} (D_S + 3 \lambda_S) \sin \theta \Big]^2 , \no \\
A_{\Lambda^*(1670) \, \eta} &=& \frac{2}{3} \Big[ D_S \cos \theta +
    \sqrt{2} \frac{F_\pi}{F_0} (D_S + 3 \lambda_S) \sin \theta  \Big]^2    
\eeqa
where we have taken $\eta$-$\eta'$ mixing into account.
For $F_0$ we employ the large $N_c$ value $F_0 = F_\pi$.
Using the experimental values for the decay widths
we arrive at the central values
\beqa
D_P &=& 0.32 ,  \qq  F_P = 0.16  \no \\
D_S &=& 0.37 ,  \qq  F_S = -0.21  ,  \qq  \lambda_S = -0.07
\eeqa
where we have chosen the signs in accordance
with the data for $\eta$ and $\eta'$ photoproduction.
We do not present the uncertainties in these parameters here, since
for the purpose of our considerations a rough estimate of these constants
is sufficient.

We now turn to the determination of the couplings $d_P, f_P$ and $d_S, f_S$ 
appearing in the electromagnetic part of the effective
resonance-ground state Lagrangian.
The decays listed in the particle data book,
which determine the coupling constants $d_S$ and $f_S$,
are $N^*(1535) \rightarrow N \gamma$, see also \cite{B2}.
The width is given by 
\beq  \label{dec}
\Gamma^{ji} = \frac{1}{8 \pi M_S^2} |{\bf k}_\gamma| |{\cal T}^{ji}|^2
\eeq
with
\beq  \label{ene}
|{\bf k}_\gamma| = E_\gamma = \frac{1}{2 M_S} (M_S^2- M_B^2) 
\eeq
being the three-momentum of the photon 
in the rest frame of the resonance.
For the transition matrix one obtains
\beq
|{\cal T}^{ji}|^2 = \, 128 \, e^2 \, ( p_i \cdot k)^2 \, (C^{ji})^2
\eeq
with $p_i$ the momentum of the decaying baryon and the coefficients
\beq
C^{p^*(1535) \, p } = \frac{1}{3} \, d_S + \,  f_S  , \qq
C^{n^*(1535) \, n} = - \frac{2}{3} \, d_S   \qq .
\eeq
Using the experimental values for the decay widths
we arrive at the central values
\beq
d_S  = -0.07 \: \mbox{GeV}^{-1}  , \qq 
f_S  =   - 0.06 \: \mbox{GeV}^{-1}  .
\eeq
For the determination of $d_P$ and $f_P$ we use the decays
$N^*(1440) \rightarrow N \gamma$.
One has to replace the resonance mass by $M_{P} \simeq 1440$ MeV 
in Eq.~(\ref{dec},\ref{ene}) and the coefficients read
\beq
C^{p^*(1440) \, p} = \, \frac{1}{3} \, d_P + \, f_P   , \qq
C^{n^*(1440) \, n} = - \frac{2}{3} \,  d_P   .
\eeq
The fit to the decay widths delivers
\beq
d_P = - 0.05 \: \mbox{GeV}^{-1}   , \qq 
f_P = 0.08 \:\mbox{GeV}^{-1}  .
\eeq

\newpage

\section*{Table captions}

\begin{enumerate}

\item[Table 1] Given are the multipoles for $\eta$ photoproduction off the
               proton at c.m. energy $s=708$ MeV both with and without 
               resonances in units of 10$^{-3}/m_{\pi^+}$. 
               The first row denotes the contributions from only 
               the Born terms. In the following rows we have added vector
               mesons, spin-1/2$^+$
               and spin-1/2$^-$ baryon resonances, respectively. The last row
               gives our final result, including all resonances considered in
               this work.

\end{enumerate}

\vskip 0.4in

%%%% figure captions

\section*{Figure captions}

\begin{enumerate}

\item[Fig.1] Shown are the Born terms for photoproduction on the proton.
             The photon is given by a wavy line. Solid and dashed lines denote
             proton and pseudoscalar mesons, respectively.  

\item[Fig.2] Vector meson exchange.  The photon is given by a wavy line.
             Solid and dashed lines denote nucleons and pseudoscalar mesons, 
             respectively. The double line represents the vector meson. 

\item[Fig.3] Baryon resonance contributions.  
             The photon is given by a wavy line.
             Solid and dashed lines denote nucleons and pseudoscalar mesons, 
             respectively. The double line represents the baryon resonances
             $P_{11}(1440)$ or $S_{11}(1535)$. 

\item[Fig.4] Total cross sections for $\gamma p \rightarrow p \eta$ (a),
             $\gamma p \rightarrow p \eta'$ (b),
             $\gamma n \rightarrow n \eta$ (c) and
             $\gamma n \rightarrow n \eta'$ (d).
             The dashed lines denote the contributions from the Born terms,
             the full lines are our results including the resonances.
             The reaction $\gamma p \rightarrow p \eta$ in (a) is compared with
             data from \cite{Kru}. We do not show data from \cite{Pl} for
             $\eta'$ photoproduction since the energy bins of this experiment
             are 100 and 200 MeV wide.

\item[Fig.5] a) Differential cross section of $\gamma p \rightarrow p \eta$
             for the photon energy $E_\gamma = 716$ MeV which is compared with
             the data from \cite{Kru}.\\
             b) Differential cross section for $\gamma p \rightarrow p \eta'$
             at $E_\gamma = 1.45$ GeV. The data are the energy-integrated
             angular distributions for 1.44 GeV $< E_\gamma <$ 1.54 GeV 
             \cite{Pl}.

\item[Fig.6] Shown are the multipoles $E_{0+}$ (a), $M_{1+}$ (b), 
             $M_{1-}$ (c) and $E_{1+}$ (d) for $\eta$ photoproduction off the
             nucleons. The full and dashed lines denote
             $\gamma p \rightarrow p \eta$ and  $\gamma n \rightarrow n \eta$,
             respectively.

\item[Fig.7] Shown are the multipoles $E_{0+}$ (a), $M_{1+}$ (b), 
             $M_{1-}$ (c) and $E_{1+}$ (d) for $\eta'$ photoproduction off the
             nucleons. The full and dashed lines denote
             $\gamma p \rightarrow p \eta'$ and  
             $\gamma n \rightarrow n \eta'$, respectively.

\item[Fig.8] Energy dependence of the coefficients A (a), B (b), C (c)
             from Eq. (\ref{coeff}) for $\eta$ photoproduction.
             The full and dashed lines denote
             $\gamma p \rightarrow p \eta$ and  $\gamma n \rightarrow n \eta$,
             respectively.

\item[Fig.9] Energy dependence of the coefficients A (a), B (b), C (c)
             from Eq. (\ref{coeff}) for $\eta'$ photoproduction.
             The full and dashed lines denote
             $\gamma p \rightarrow p \eta'$ and  
             $\gamma n \rightarrow n \eta'$, respectively.

\end{enumerate}

\newpage

%%%%%%% Tables   %%%%%%
\begin{center}

\begin{table}[bht]  \label{tab1}
\begin{center}
 % \medskip 
\begin{tabular}{l|c|c|c|c}
  & $E_{0+}$ & $M_{1+}$ & $M_{1-}$ & $E_{1+}$ \\
\hline
$ N $  & -4.47 & 0.073 & -0.10 & 0.009  \\
$ N, V $  & 2.48 & 0.084 & 0.27 & -0.006  \\
$ N, P_{11} $  & -4.40 & 0.079 & -0.32 & 0.009  \\
$ N, S_{11} $  & 6.36 & 0.076 & -0.09 & 0.008  \\
$ N, V, P_{11}, S_{11} $  & 13.39 & 0.092 & 0.06 
         & -0.007  \\
\hline
\end{tabular}
\end{center}
\end{table}
\vskip 0.7cm

Table  1

\end{center}

\newpage

%%%%%%% Figures   %%%%%%
\begin{center}
 
\begin{figure}[bth]
\centering
%\leavemode
\centerline{
\epsfbox{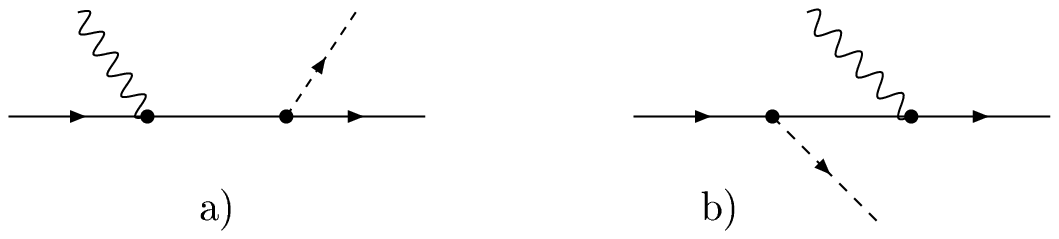}}
\end{figure}

\vskip 0.7cm

Figure 1

\vskip 2cm

\begin{figure}[tbh]
\centering
%\leavemode
\centerline{
\epsfbox{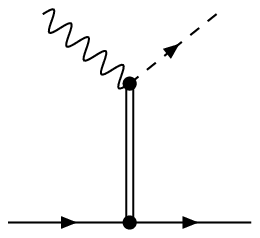}}
\end{figure}

\vskip 0.7cm

Figure 2

\vskip 2cm

\begin{figure}[tbh]
\centering
%\leavemode
\centerline{
\epsfbox{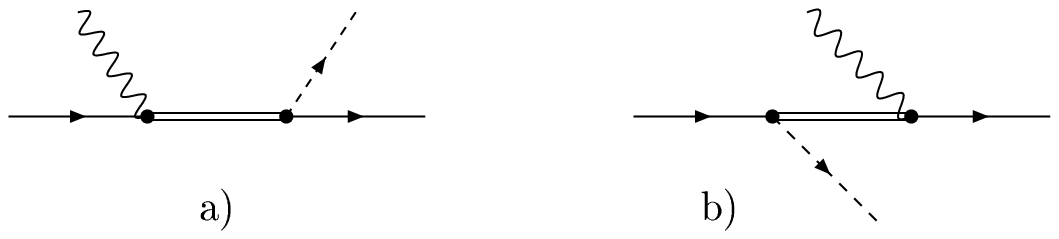}}
\end{figure}

\vskip 0.7cm

Figure 3

\begin{figure}[tbh]
\centering
%\leavemode
\begin{picture}(300,380)  
\put(-20,260){\makebox(100,120){\epsfig{file=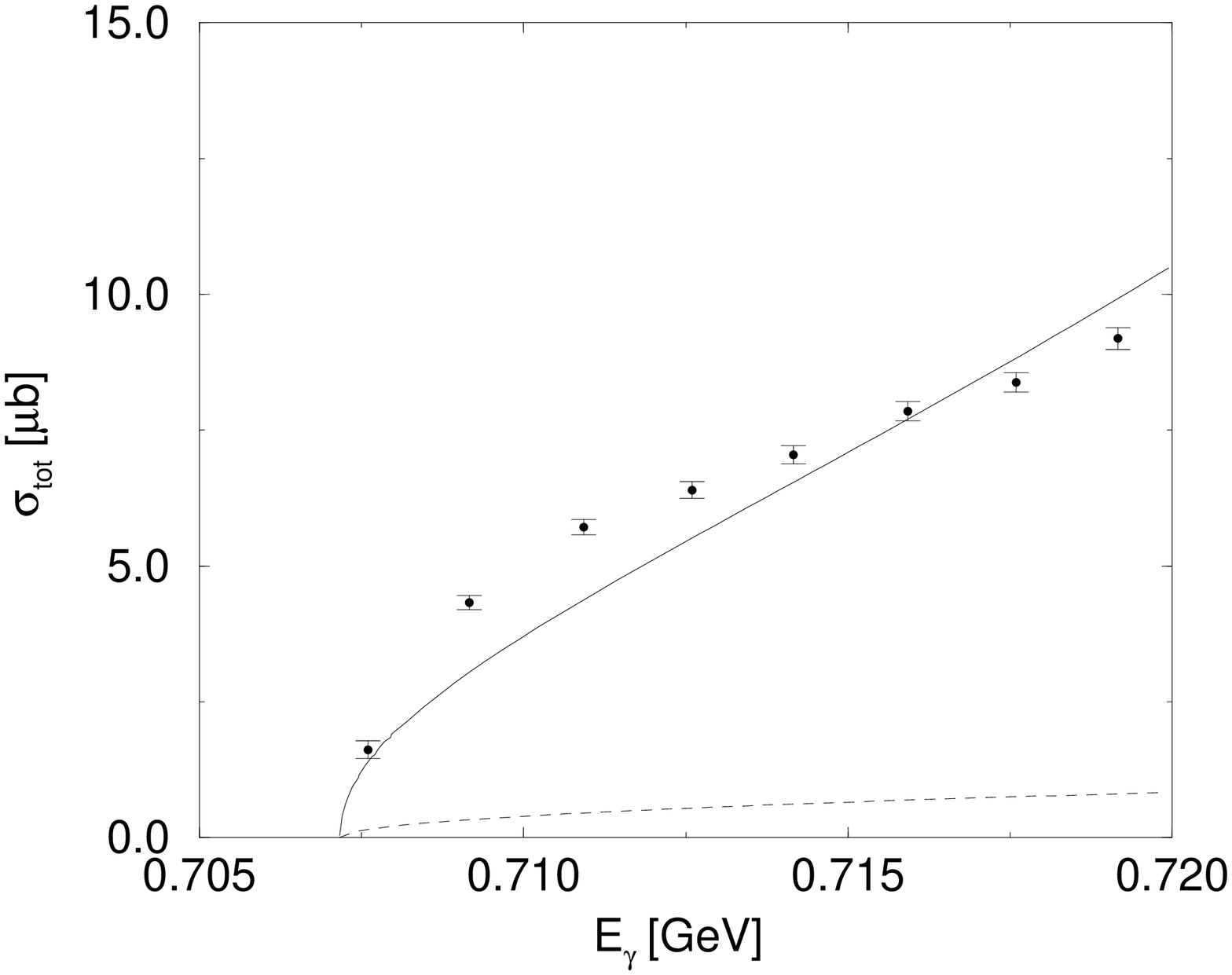,width=7.0cm,
              angle=-90}}}
\put(240,260){\makebox(100,120){\epsfig{file=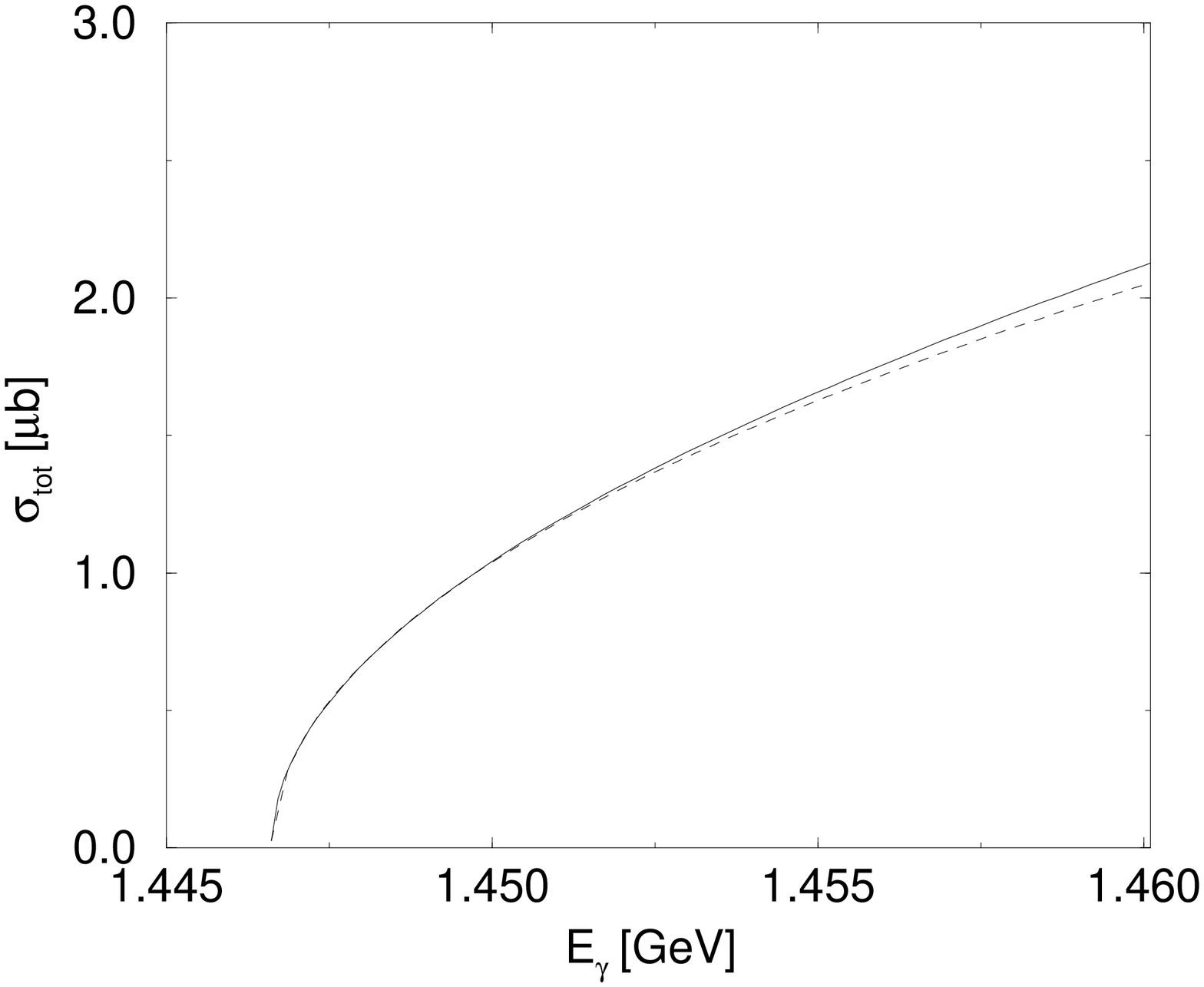,width=7.0cm,
               angle=-90}}}
\put(-20,0){\makebox(100,120){\epsfig{file=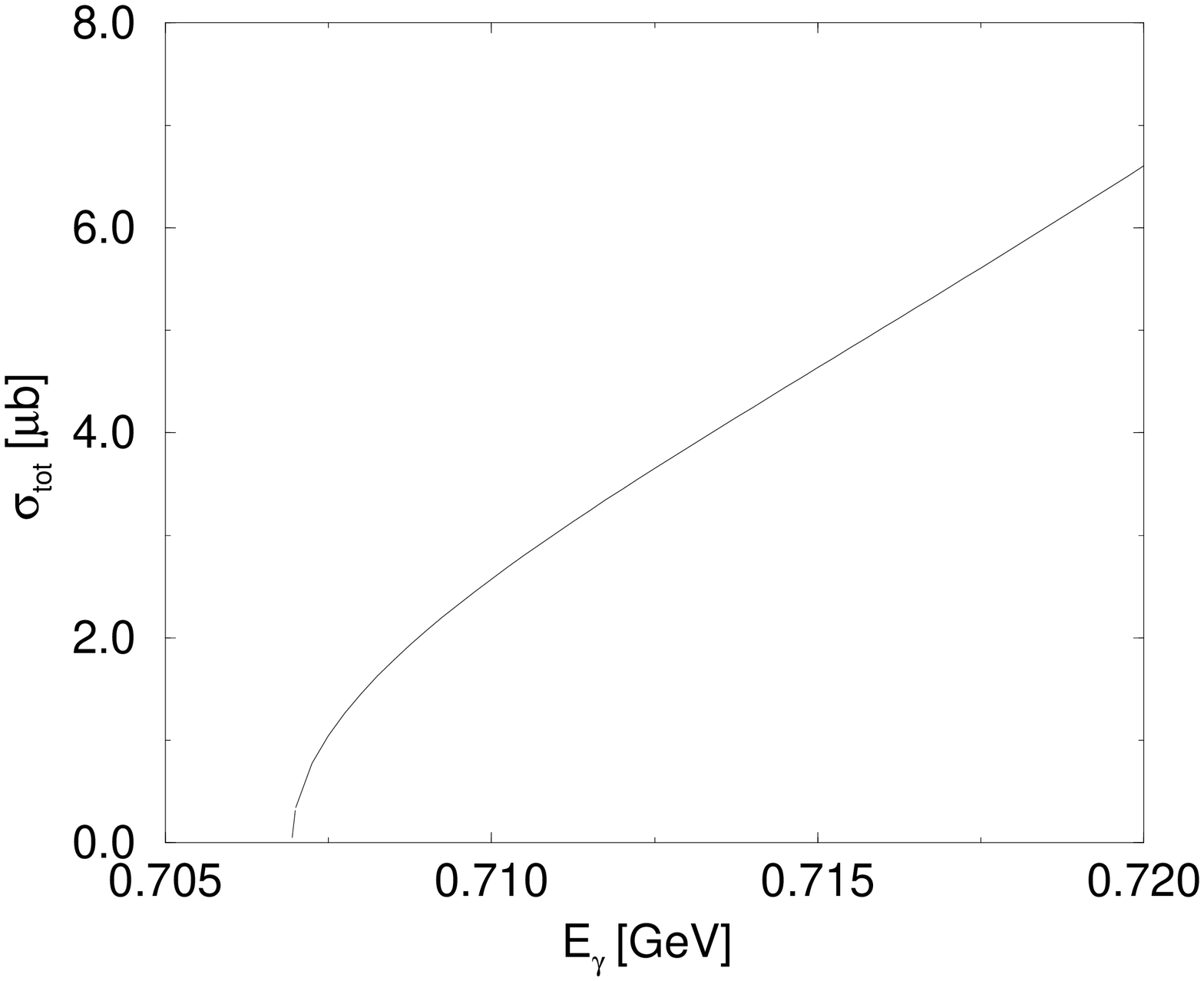,width=7.0cm,angle=-90}}}
\put(240,0){\makebox(100,120){\epsfig{file=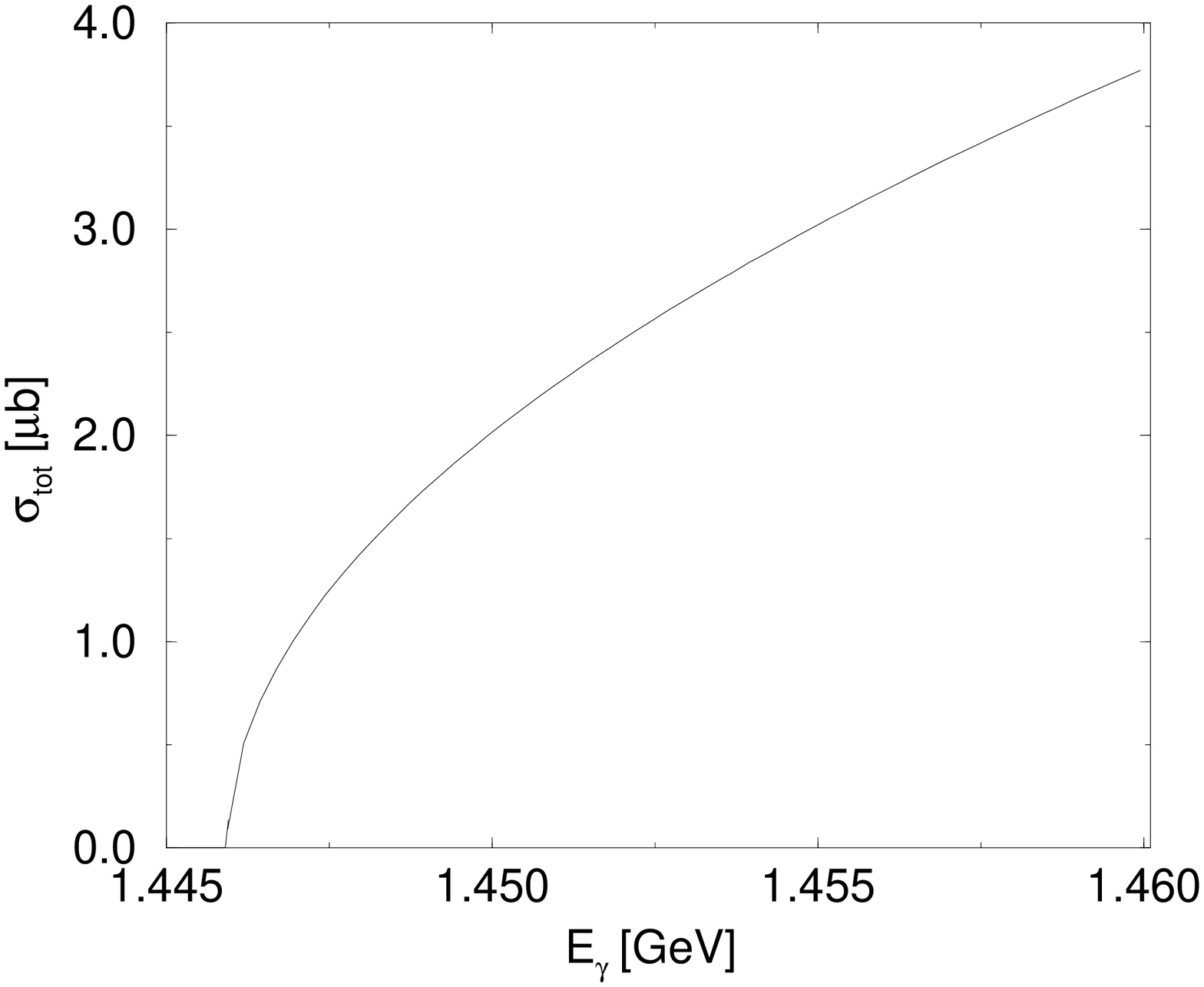,width=7.0cm,angle=-90}}}
\put(30,200){$a)$}
\put(290,200){$b)$}
\put(30,-60){$c)$}
\put(290,-60){$d)$}
\end{picture}
\vskip 2.8cm

Figure 4

\end{figure}

\begin{figure}[tbh]
\centering
%\leavemode
\begin{picture}(300,380)  
\put(-20,260){\makebox(100,120){\epsfig{file=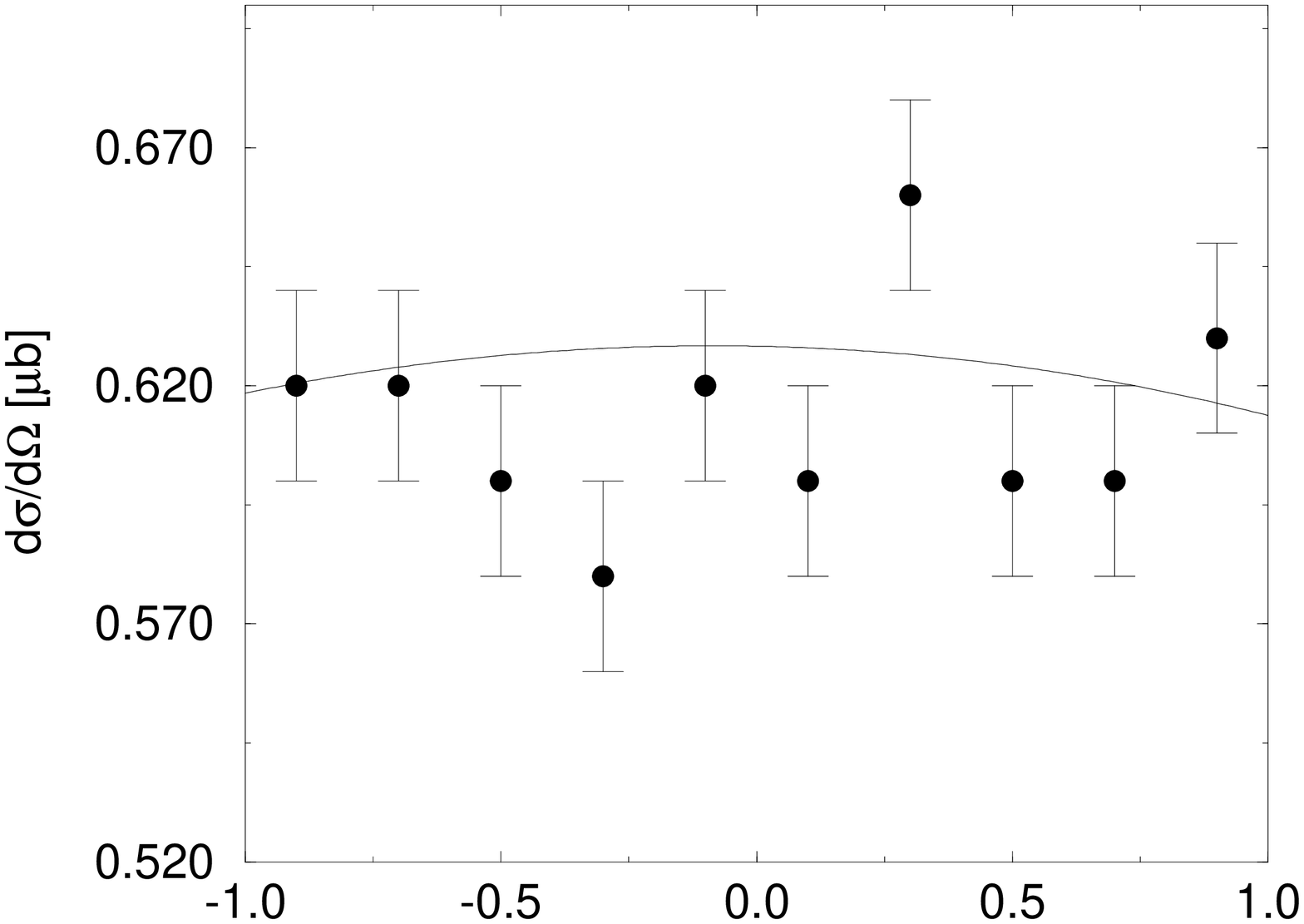,width=7.0cm,
              angle=-90}}}
\put(240,260){\makebox(100,120){\epsfig{file=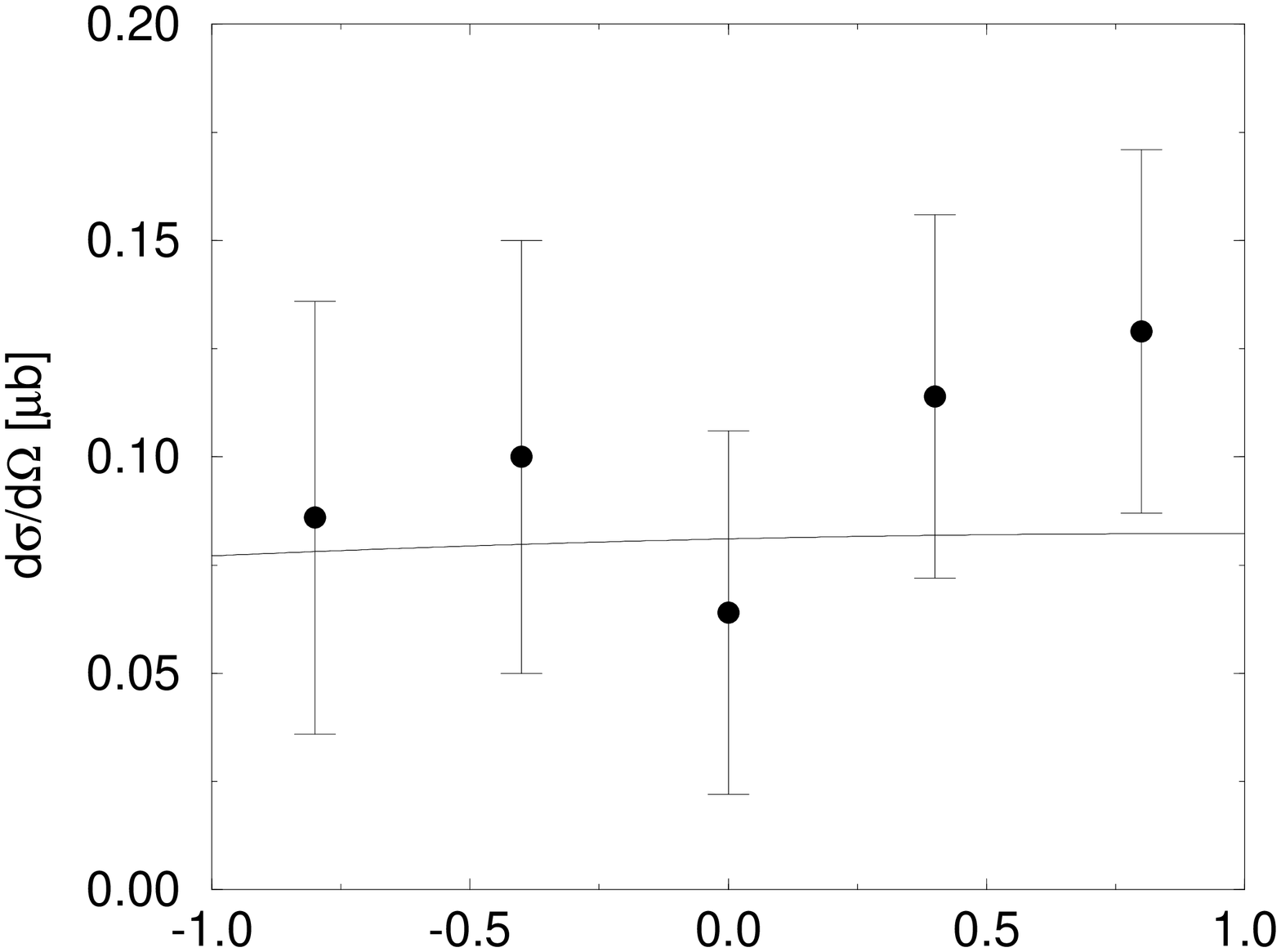,width=7.0cm,
               angle=-90}}}
\put(30,210){{\footnotesize $\cos \vartheta$}}
\put(290,210){{\footnotesize $\cos \vartheta$}}
\put(30,180){$a)$}
\put(290,180){$b)$}
\end{picture}
\vskip -4.8cm

Figure 5

\end{figure}

\begin{figure}[tbh]
\centering
%\leavemode
\begin{picture}(300,380)  
\put(-20,260){\makebox(100,120){\epsfig{file=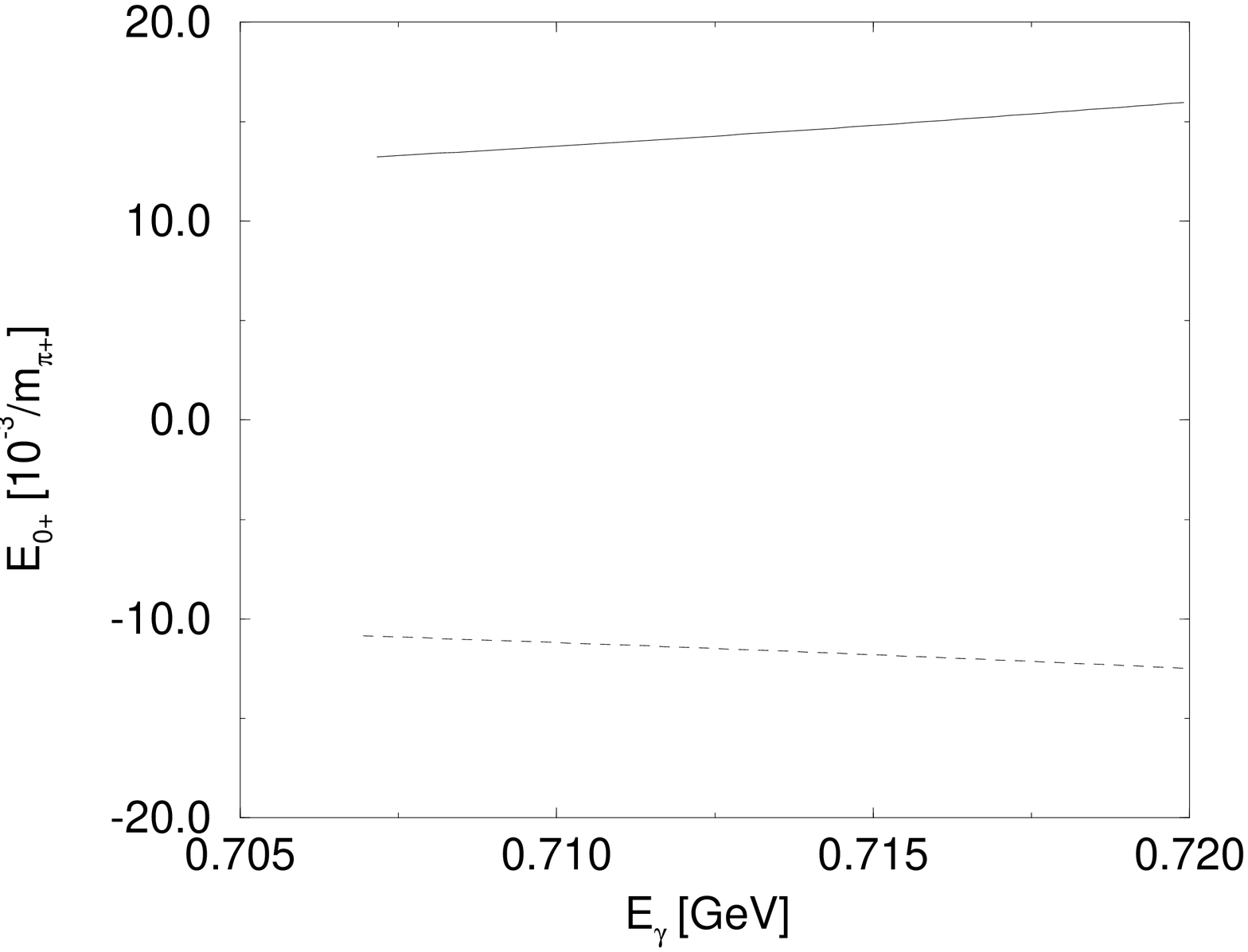,width=7.0cm,
              angle=-90}}}
\put(240,260){\makebox(100,120){\epsfig{file=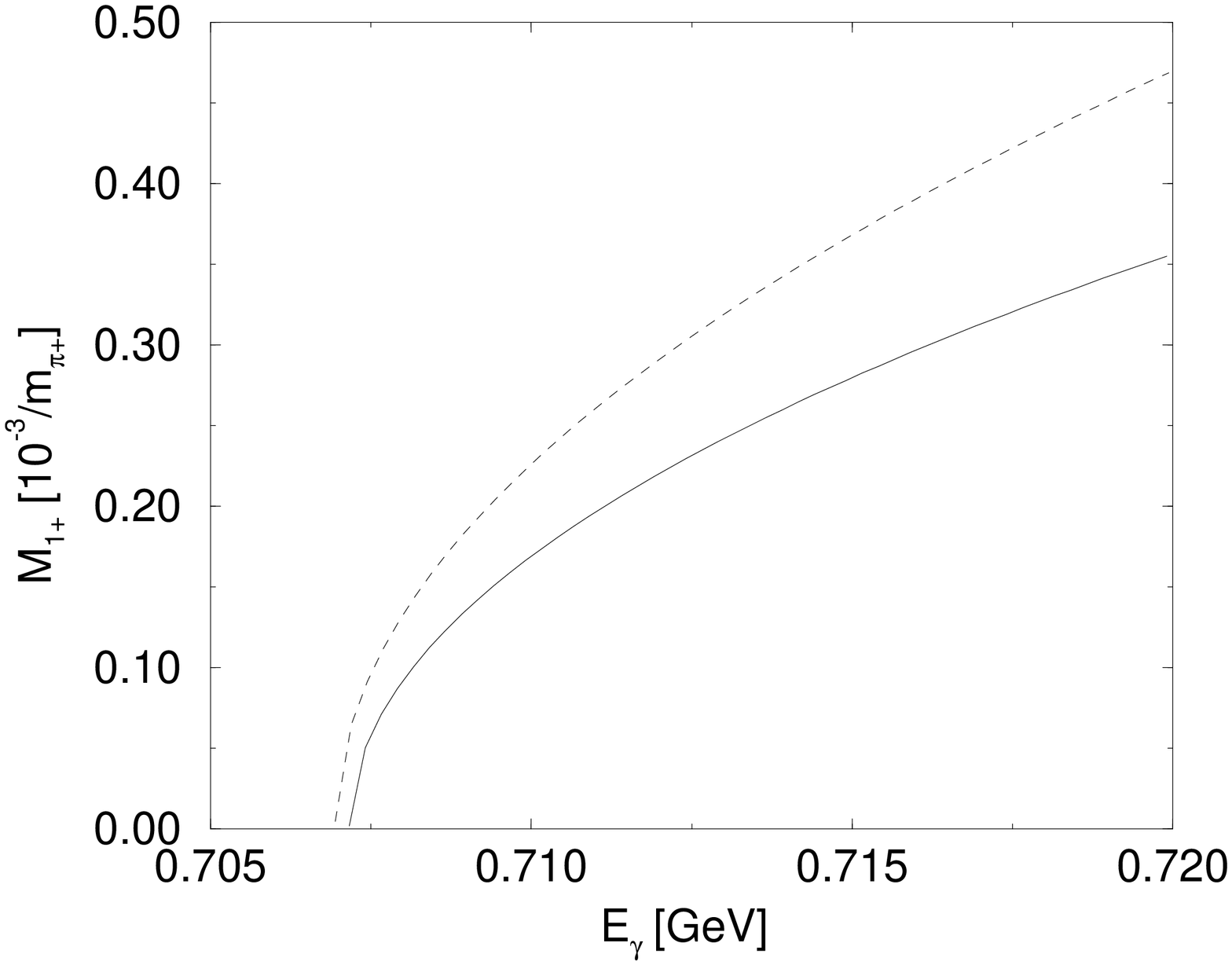,width=7.0cm,
               angle=-90}}}
\put(-20,0){\makebox(100,120){\epsfig{file=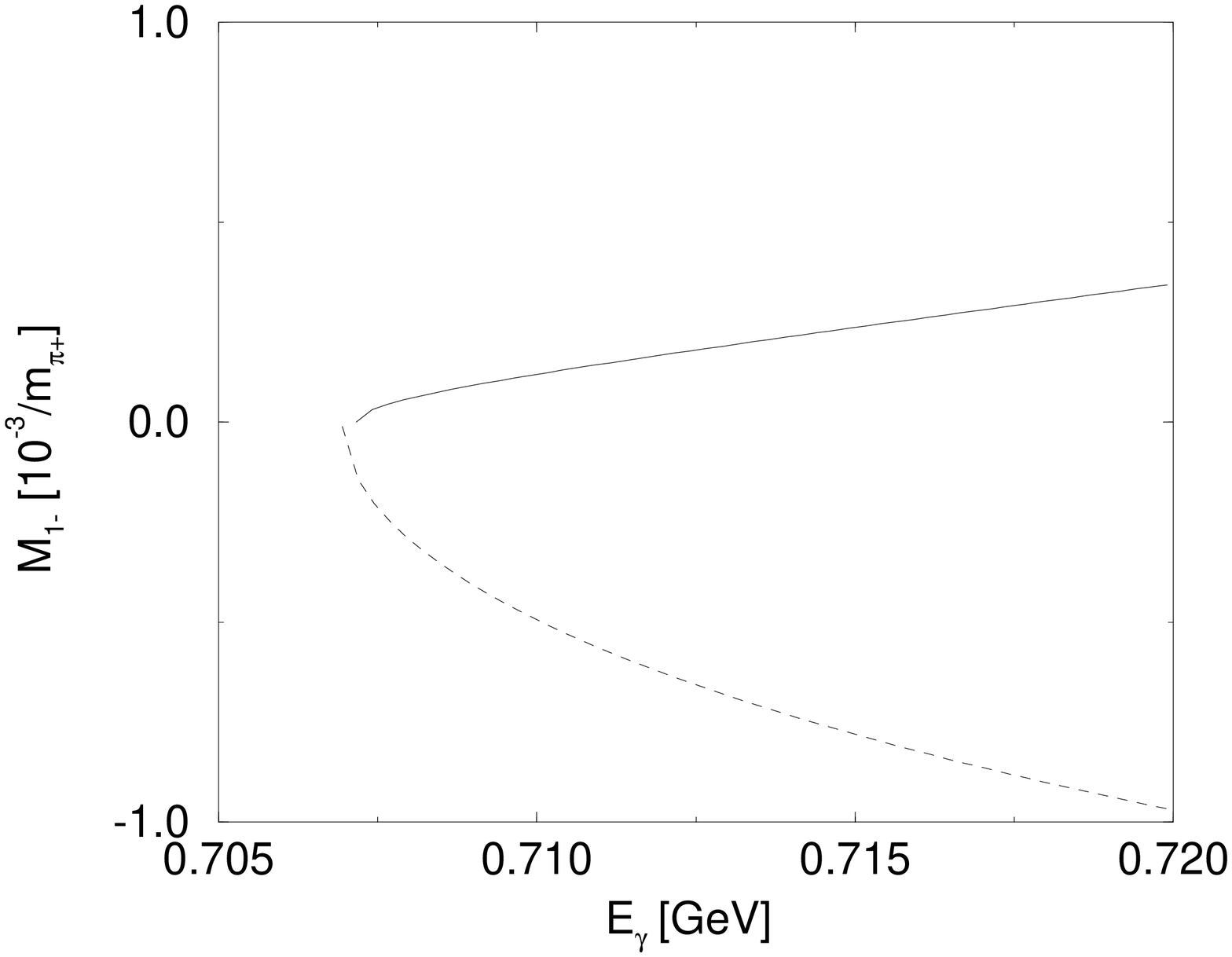,width=7.0cm,angle=-90}}}
\put(240,0){\makebox(100,120){\epsfig{file=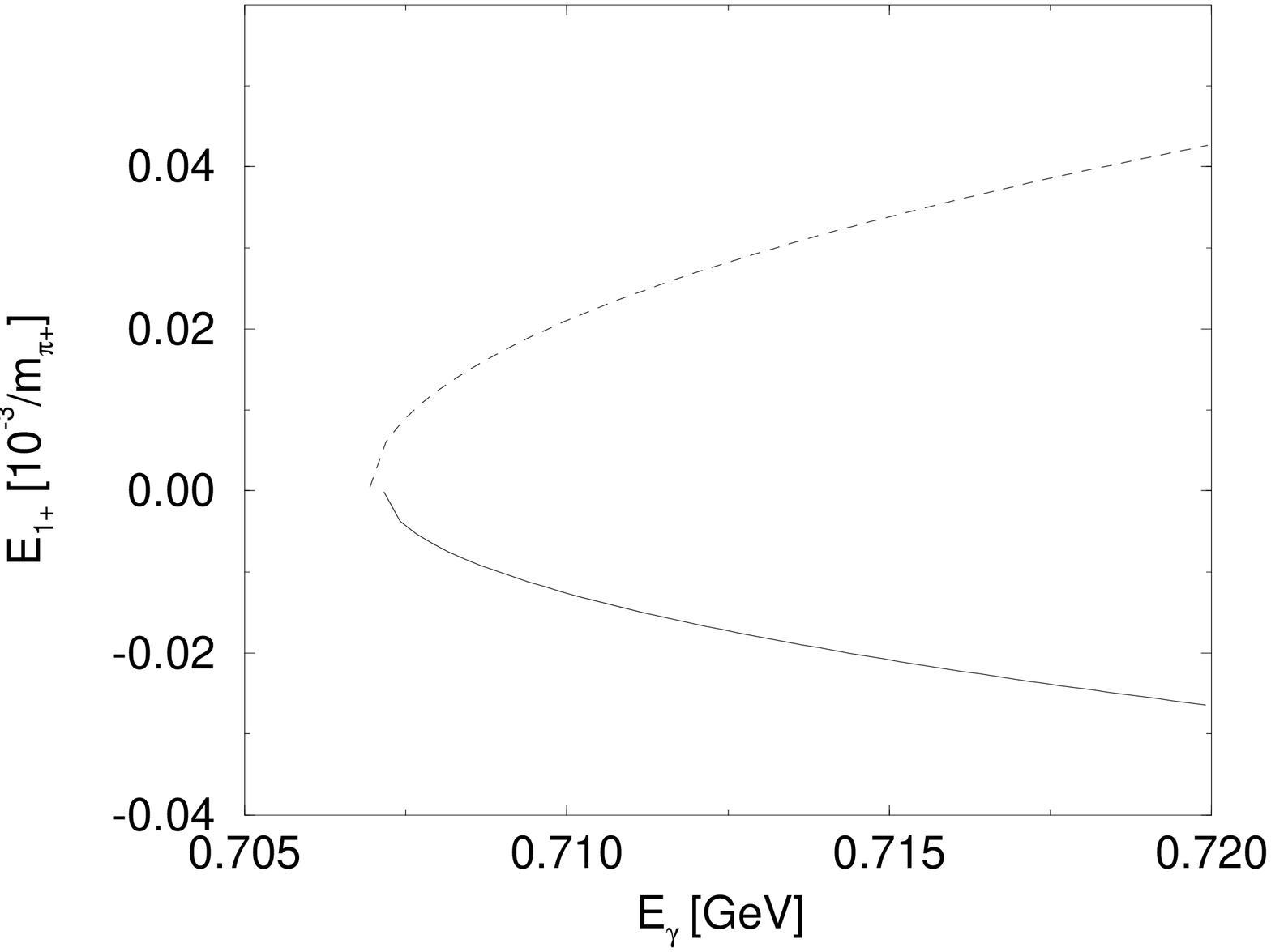,width=7.0cm,angle=-90}}}
\put(30,200){$a)$}
\put(290,200){$b)$}
\put(30,-60){$c)$}
\put(290,-60){$d)$}
\end{picture}
\vskip 2.8cm

Figure 6

\end{figure}

\begin{figure}[tbh]
\centering
%\leavemode
\begin{picture}(300,380)  
\put(-20,260){\makebox(100,120){\epsfig{file=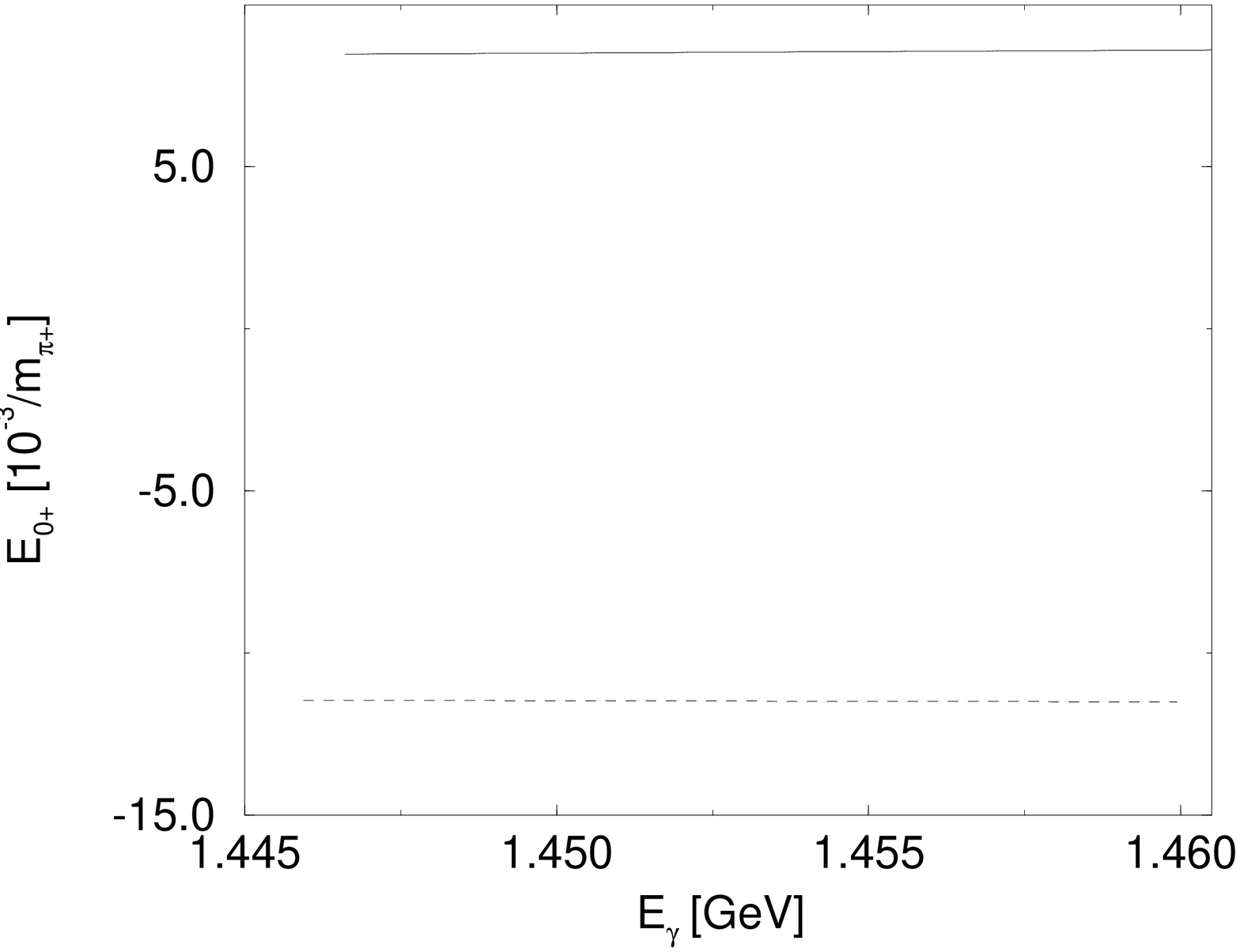,width=7.0cm,
              angle=-90}}}
\put(230,260){\makebox(100,120){\epsfig{file=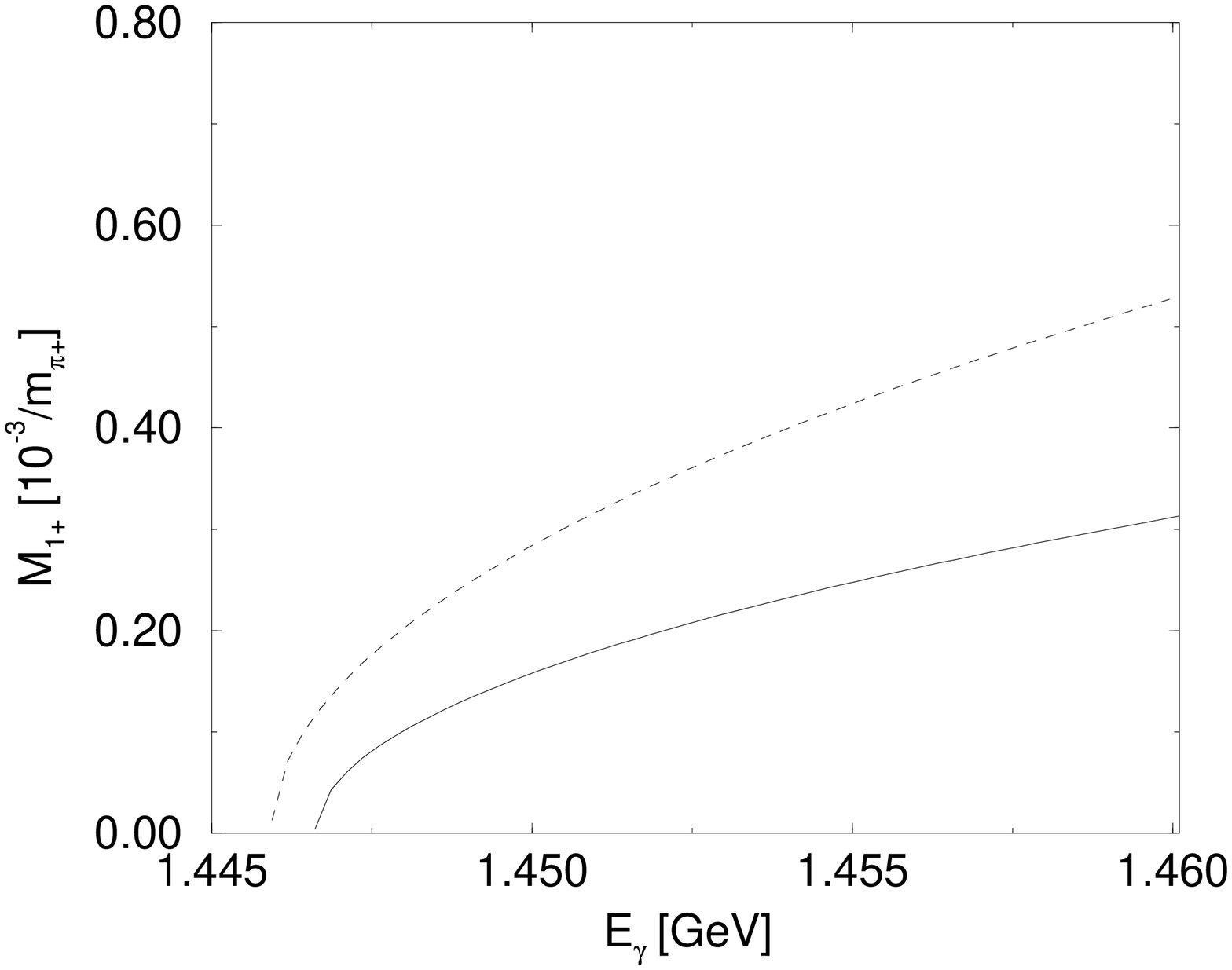,width=7.0cm,
               angle=-90}}}
\put(-20,0){\makebox(100,120){\epsfig{file=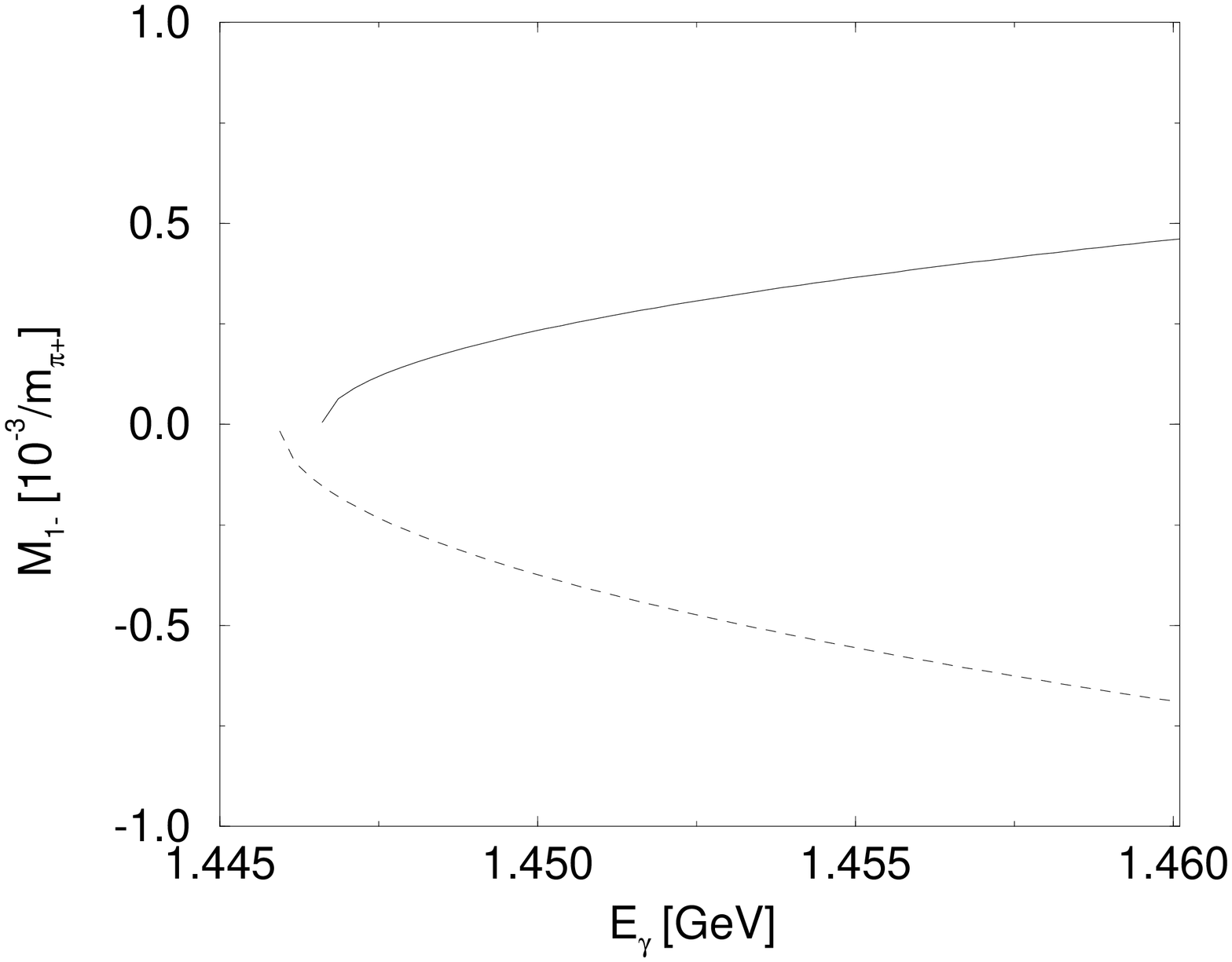,width=7.0cm,angle=-90}}}
\put(230,0){\makebox(100,120){\epsfig{file=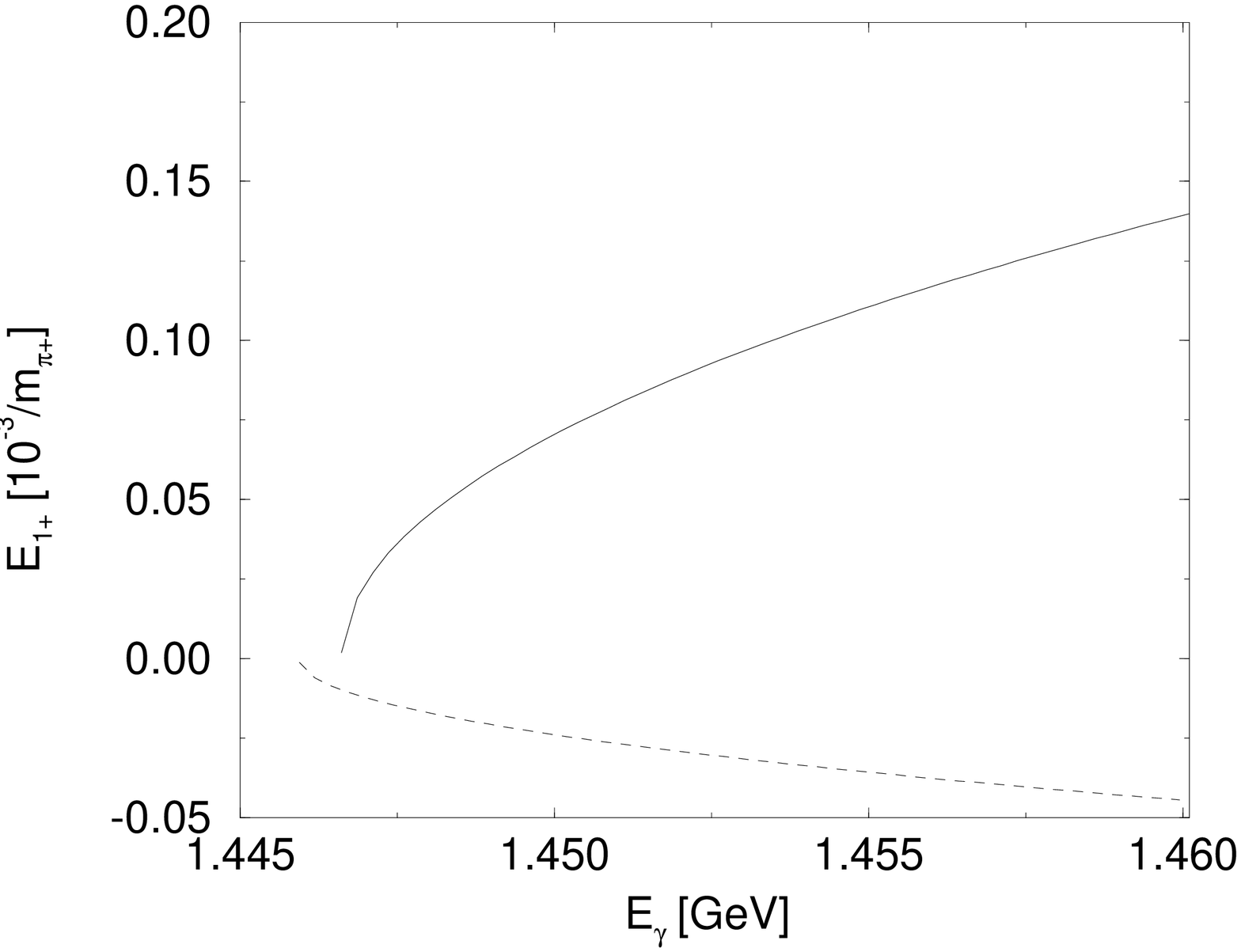,width=7.0cm,angle=-90}}}
\put(30,200){$a)$}
\put(280,200){$b)$}
\put(30,-60){$c)$}
\put(280,-60){$d)$}
\end{picture}
\vskip 2.8cm

Figure 7

\end{figure}

\begin{figure}[tbh]
\centering
%\leavemode
\begin{picture}(300,380)  
\put(-20,260){\makebox(320,120){\epsfig{file=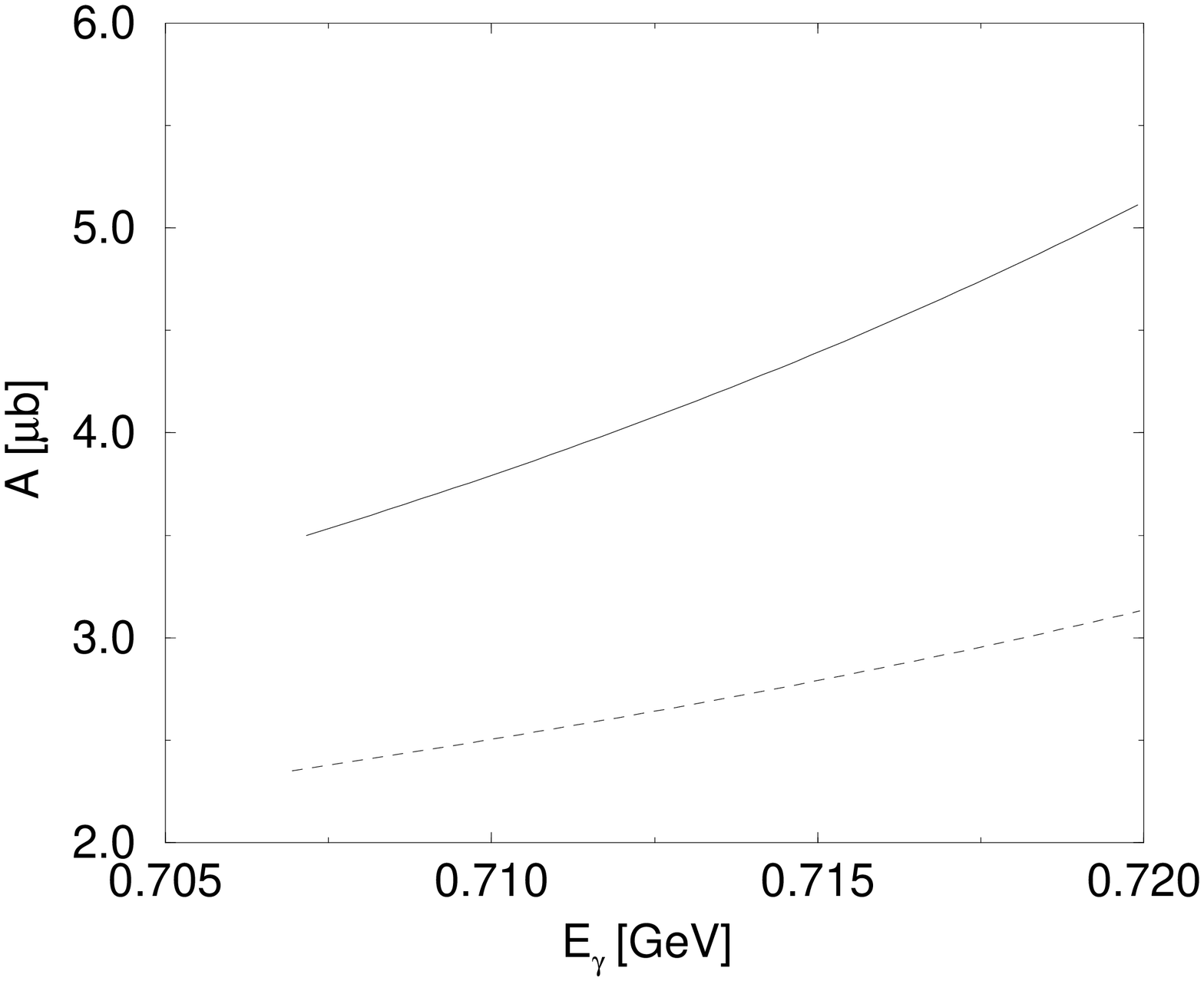,width=6.0cm,
              angle=-90}}}
\put(-20,90){\makebox(320,120){\epsfig{file=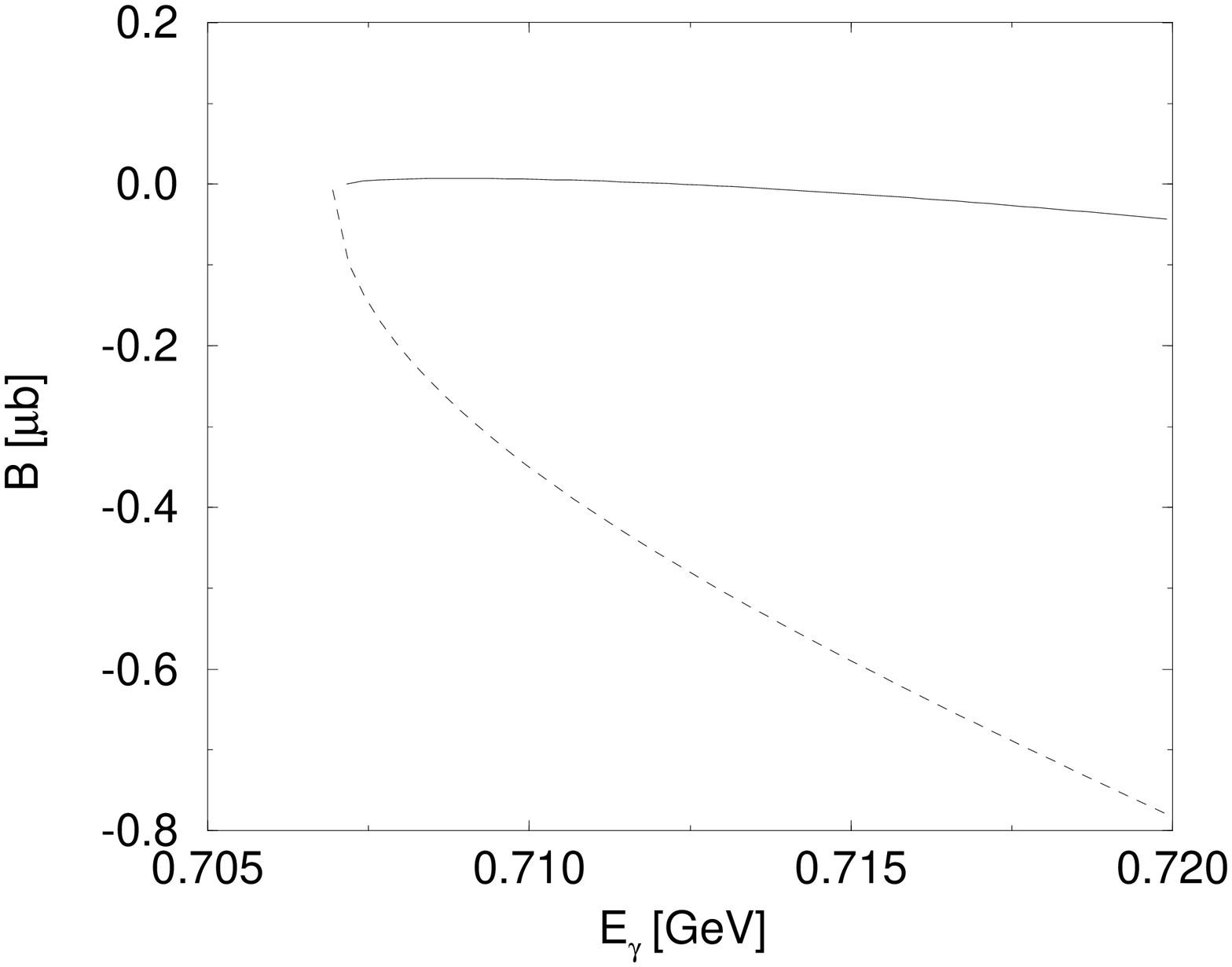,width=6.0cm,angle=-90}}}
\put(-20,-80){\makebox(320,120){\epsfig{file=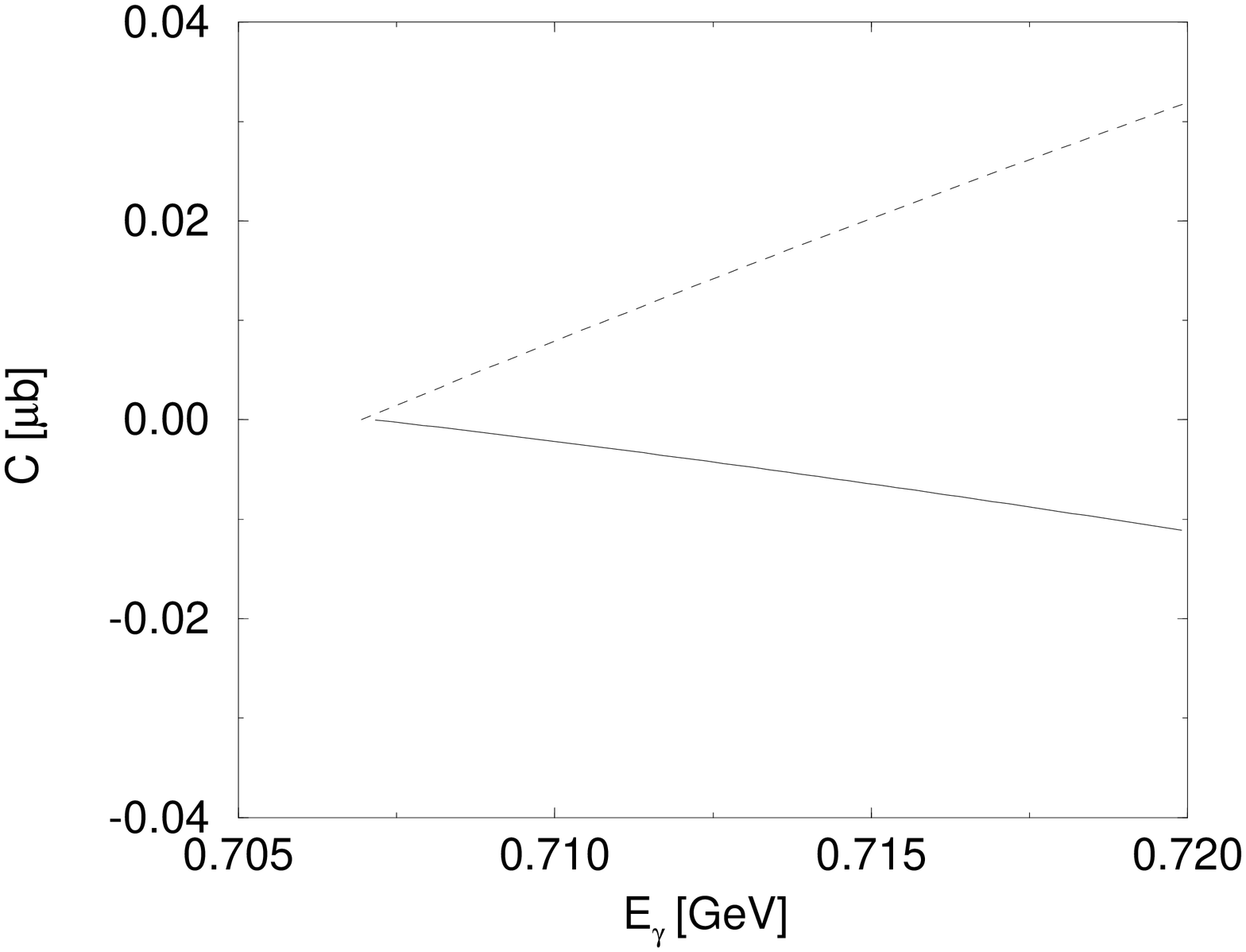,width=6.0cm,
               angle=-90}}}
\put(245,300){$a)$}
\put(245,130){$b)$}
\put(245,-40){$c)$}
\end{picture}
\vskip 4.8cm

Figure 8

\end{figure}

\begin{figure}[tbh]
\centering
%\leavemode
\begin{picture}(300,380)  
\put(-20,260){\makebox(320,120){\epsfig{file=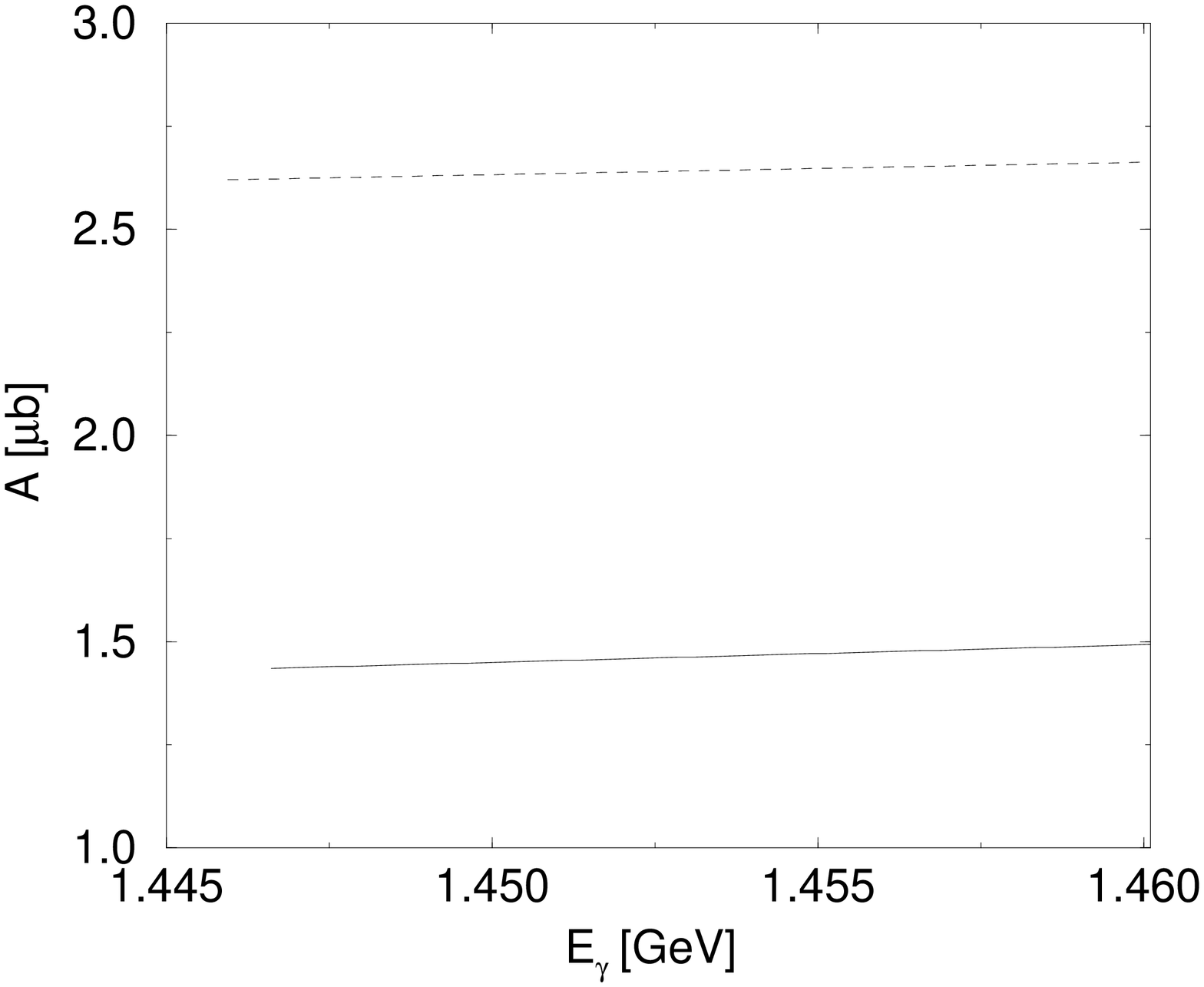,width=6.0cm,
              angle=-90}}}
\put(-20,90){\makebox(320,120){\epsfig{file=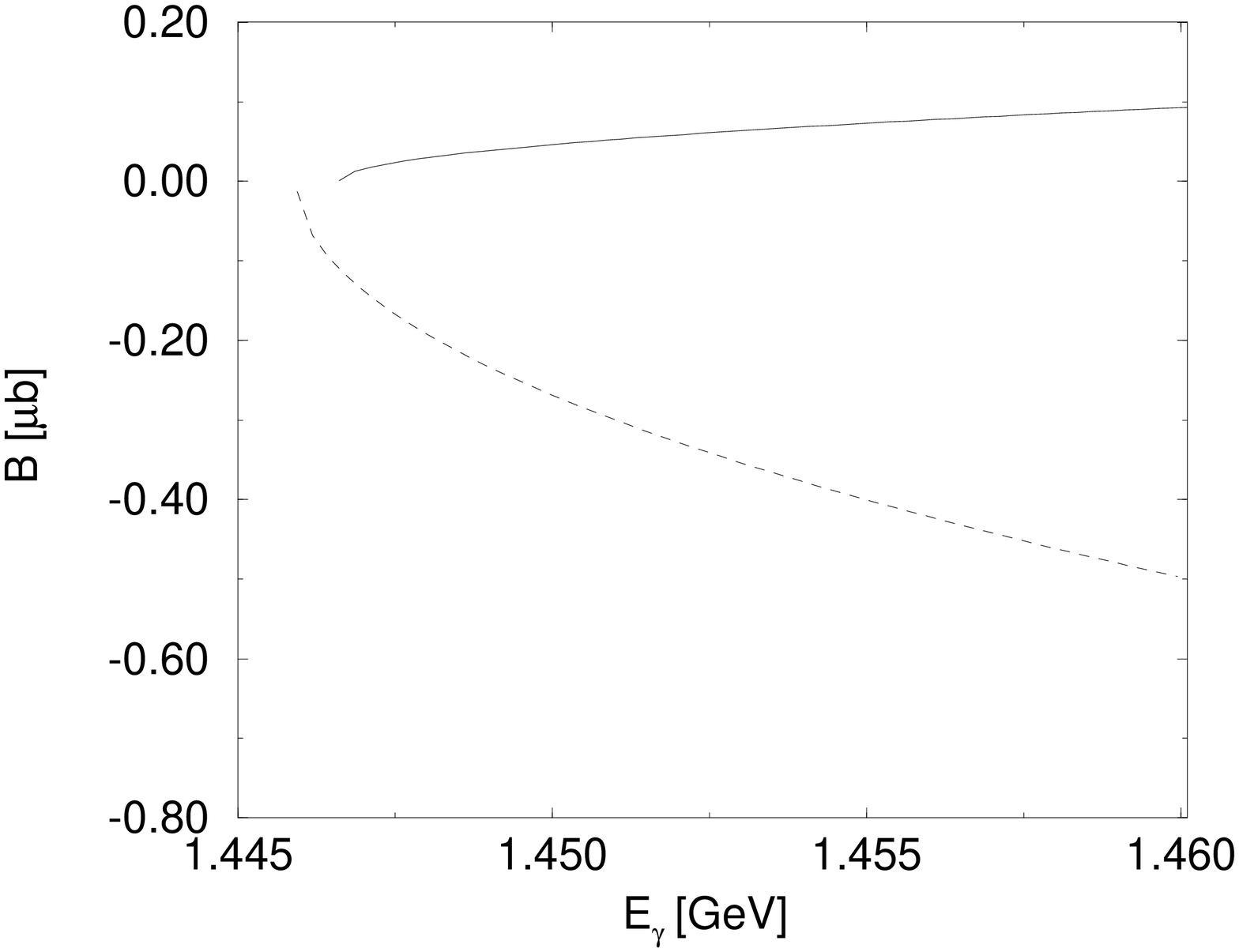,width=6.0cm,angle=-90}}}
\put(-20,-80){\makebox(320,120){\epsfig{file=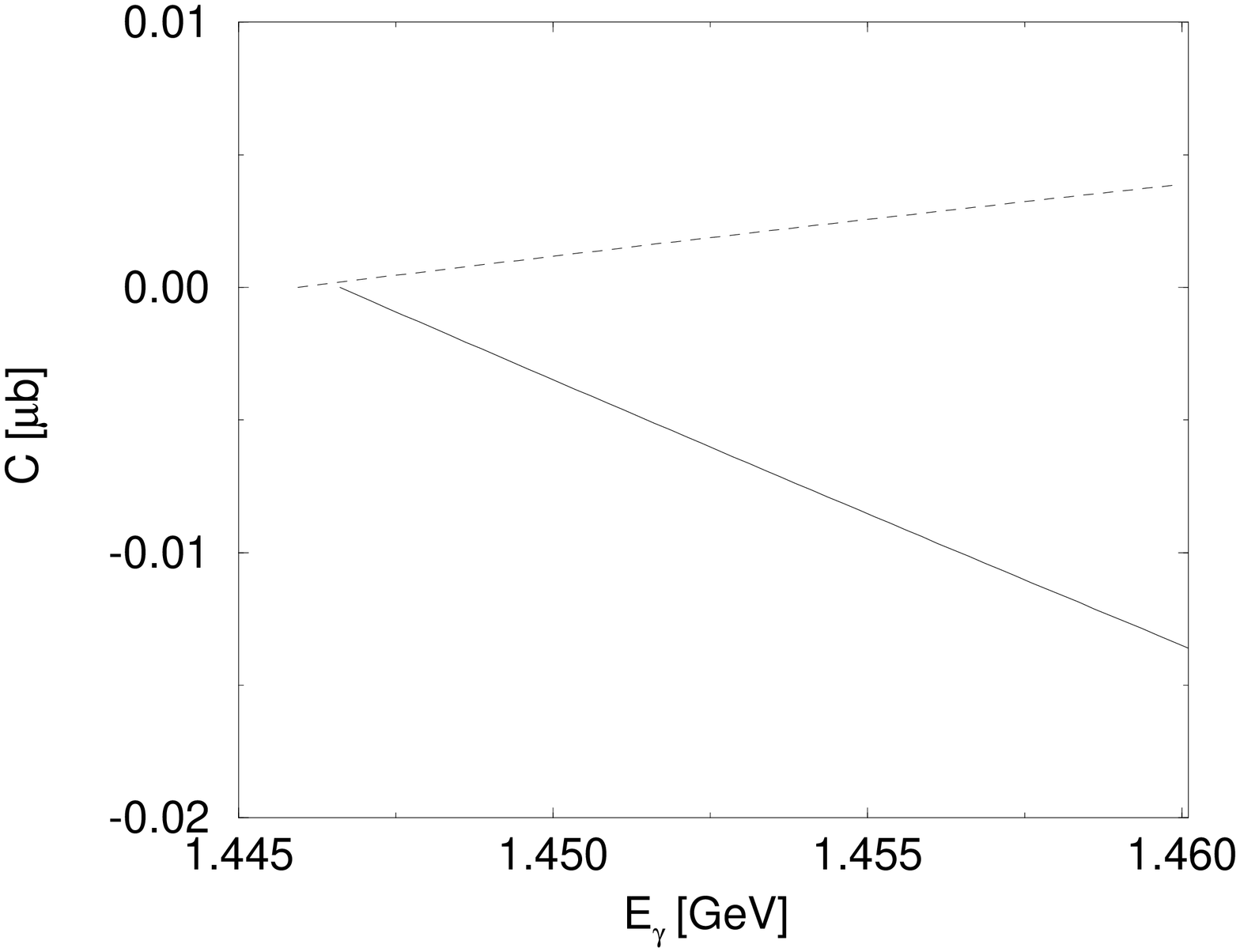,width=6.0cm,
               angle=-90}}}
\put(245,300){$a)$}
\put(245,130){$b)$}
\put(245,-40){$c)$}
\end{picture}
\vskip 4.8cm

Figure 9

\end{figure}

\end{center}

\end{document}